\documentclass[a4paper,fleqn]{cas-dc}

\usepackage[authoryear]{natbib}

\usepackage{amsmath,amssymb}
\usepackage{booktabs}

\usepackage{xcolor}
\usepackage{xurl}
\usepackage{float}

\usepackage{placeins}


\begin{document}
\let\WriteBookmarks\relax
\def\floatpagepagefraction{1}
\def\textpagefraction{.001}

\shorttitle{Shared neighbourhood pattern of reported violence}
\shortauthors{} 

\title[mode=title]{Changing contribution of a shared neighbourhood pattern to reported sexual and non-sexual violence against women and girls in Medellín, Colombia, 2018--2022}

%
\author[1]{Tomás Rodríguez~Taborda}
\cormark[1]
\ead{torodriguezt@unal.edu.co}

\author[1]{Alejandra-Estefanía Patiño-Hoyos}[orcid=0000-0001-8128-1307]
\ead{aepatino@unal.edu.co}

\author[2]{Simón Ruíz-Martínez}[orcid=0000-0002-8929-7394]
\ead{siruizm@unal.edu.co}

\author[1]{Johnatan Cardona-Jiménez}[orcid=0000-0002-6370-8837]
\ead{jcardonj@unal.edu.co}

\affiliation[1]{organization={Departamento de Estadística, Facultad de Ciencias, Universidad Nacional de Colombia},
            addressline={Carrera 65 No. 59A-110},
            city={Medellín},
            postcode={050034},
            country={Colombia}}

\affiliation[2]{organization={Departamento de Ciencia Política, Facultad de Ciencias Humanas y Económicas, Universidad Nacional de Colombia},
            addressline={Carrera 65 No. 59A-110},
            city={Medellín},
            postcode={050034},
            country={Colombia}}

\cortext[1]{Corresponding author}

\begin{abstract}
During the COVID-19 restrictions, notifications of violence against women and girls may have changed unevenly across neighbourhoods. We asked whether the contribution of a shared neighbourhood pattern changed over this period, and whether any change was common to sexual and non-sexual violence or specific to one. We analysed $46{,}390$ suspected cases notified to the public-health surveillance system of Medellín, Colombia, across 285 neighbourhoods in 2018--2022. A Bayesian shared-component model with age-standardised expected counts allowed the loading of a common spatial field on each form of violence to differ between the pre-pandemic (2018--2019), restriction (2020--2021) and post-restriction (2022) periods. Across the urban communes, the shared field was generally higher where living conditions were poorer. During the restrictions, its loading increased by $0.26$ (95\% credible interval $0.16$--$0.37$) for non-sexual violence and by $0.23$ ($0.09$--$0.37$) for sexual violence, without clear evidence of a change in their ratio ($-0.06$, $-0.21$ to $0.10$). Between neighbourhoods at the 90th and 10th percentiles of the shared field, the shared component's contribution to their risk ratio widened by about $36\%$ for non-sexual and $31\%$ for sexual violence. In 2022, the non-sexual loading remained $0.19$ ($0.07$--$0.31$) above its pre-pandemic value, whereas the sexual-violence loading was estimated close to its pre-pandemic level. The direction of these changes was retained among notifications from health institutions only, excluding family commissaries whose reporting changed during the period, and under alternative specifications. Twenty-five neighbourhoods had annual joint probabilities above $0.95$ of both relative risks exceeding one in all five years. The contribution of the shared neighbourhood pattern increased for both outcomes during the restrictions and remained elevated in 2022 for non-sexual violence.
\end{abstract}

\begin{highlights}
\item Sexual and non-sexual violence notifications are modelled jointly in Medellín.
\item Notifications of both forms share a geography that is higher in poorer communes.
\item In 2020--2021, notification gaps between high- and low-risk areas widened for both.
\item In 2022 the gap was still wider than before the pandemic for non-sexual violence.
\end{highlights}


\maketitle


\section{Introduction}\label{sec:introduction}

According to the World Health Organization (WHO), roughly one in every three women has ``been subjected to physical and/or sexual violence from any current or former husband or male intimate partner, or to sexual violence from someone other than a partner, or to both these forms of violence at least once in their lifetime'' \citealp[xviii]{world_health_organization_violence_2023}. In Colombia, WHO estimates indicate that 29.8\% of ever-partnered women aged 15--49 have experienced physical and/or sexual intimate partner violence during their lifetime \citep{world_health_organization_violence_2023}. Violence against women (VAW) has profound consequences for women's physical, psychological, sexual, and economic well-being \citep{WHOVAWFactsheet}. Its occurrence has been linked to structural gender inequalities \citep{Yodanis2004}, sociocultural norms that tolerate aggression \citep{Piedalueetal2020}, historical processes of violence \citep{Young2005}, and differences in institutional capacity for prevention and response \citep{GouldAgnich2016}. Because these determinants are unevenly distributed across places, VAW should then be understood as a patterned phenomenon that reaches social, institutional and territorial contexts \citep{WHO2002,EllsbergHeise2005,RussoPirlott2006}.

In Colombia, information on VAW is generated by judicial, forensic, survey-based, and health surveillance systems. Among these, the National Public Health Surveillance System (SIVIGILA), coordinated by the National Institute of Health (INS), provides continuous nationwide surveillance of suspected violence-related events \citep{INSProtocolo}. In Medellín, notifications come from both health institutions and municipal family commissaries (Comisarías de Familia). Although this system enables timely monitoring and broad territorial coverage, notifications remain subject to underreporting, differential access to services, and spatial variation in reporting practices \citep{INSInforme2023}. The resulting maps therefore describe notifications shaped by both the contexts in which violence is experienced and the opportunities to reach services and have an event recorded.

In Medellín, previous research has documented marked spatial differences in household quality of life and in the factors associated with it \citep{Sepulveda2019}. Research using mobile phone data has also identified pronounced socioeconomic segregation across neighbourhoods. Daily mobility was found to increase social mixing, particularly through the movements of residents from lower-income neighbourhoods, who travelled longer distances and visited more diverse destinations \citep{MOYAGOMEZ2021}.

During the COVID-19 lockdowns, stay-at-home orders increased the time many women spent in domestic settings with potential aggressors while disrupting support and reporting services. Previous studies and reviews reported increases in domestic and intimate partner violence during confinement worldwide \citep{Evans2020,Piquero2021}, although administrative and health-system records are difficult to interpret because lockdowns also altered access to reporting \citep{Evans2020}. Most of this evidence concerns changes in the overall level of violence. Evidence from Bogot\'a, Medell\'in and Cali shows that lockdowns reduced travel distances and altered mobility patterns differently across socioeconomic strata \citep{Valganon2023}. Such changes may also have reshaped the geography of violence, potentially changing where different forms of violence became concentrated and how strongly a common spatial pattern contributed to their notifications.

International research has documented non-random spatial patterns of violence at neighbourhood, municipal, and regional scales \citep{FoxBenson2006,Caetano2010,Beyer2015,Gracia2018,Muldoon2019}. Bayesian spatial and spatio-temporal models allow information to be shared across neighbouring areas, stabilising estimates where counts are sparse while explicitly accounting for uncertainty \citep{VicenteSER2020,VicenteJRSSA2020,Habyarimana2021}. Multivariate formulations further allow related outcomes to be analysed jointly by separating shared from outcome-specific spatial structure \citep{GelfandVounatsou2003,MartinezBeneito2013}. Shared-component models formalise this idea through latent common structures scaled by outcome-specific loadings \citep{Held2005,KnorrHeldBest2001}. More recently, \citet{Retegui2024} allowed the scaling of a shared spatio-temporal interaction to vary across periods, so that its relative contribution to the outcomes can evolve over time. In Colombia, however, Bayesian applications have largely focused on single forms of violence and relatively broad administrative units \citep{Camargo2022}.

This study examines whether the contribution of a shared neighbourhood pattern to sexual and non-sexual violence notifications among women and girls in Medellín changed during the COVID-19 restrictions. We analyse neighbourhood-level surveillance data from 2018--2022 using a bivariate spatio-temporal shared-component model to compare the pre-pandemic, restriction and post-restriction periods. The model allows us to distinguish a strengthening or weakening of the shared contribution common to both outcomes from a change in its relative contribution to sexual versus non-sexual violence.

\section{Materials and Methods}\label{sec:methods}

\subsection{Data and study population}\label{sec:data}

We used neighbourhood-level surveillance data on violence against women and girls in Medellín, Colombia, for 2018--2022. The data were obtained from MEData, the municipal open-data platform that consolidates notifications of gender-based and intrafamilial violence originating in the national public-health surveillance system (SIVIGILA) \citep{MEDataViolence}. Cases are reported using the standardised INS~875 notification form \citep{INS875}, which records demographic, contextual, and clinical information on the victim, the event, and the aggressor. In accordance with the surveillance protocol \citep{INSProtocolo}, cases enter the system as suspected and reflect contact with health or protection services rather than legal outcomes. In Medellín, notifications come from both health institutions and municipal family commissaries (Comisarías de Familia); the latter accounted for $60\%$--$71\%$ of non-sexual notifications in each year and almost no sexual violence notifications.

The study population comprised suspected cases involving female victims. Of the $53{,}490$ records in the cleaned MEData file, $49{,}693$ were notified in 2018--2022, of which $48{,}964$ concerned female victims. We excluded $2{,}572$ ($5.3\%$) without a usable neighbourhood code and 2 with a code absent from the cartography, leaving $46{,}390$ notifications. No included record duplicated another in all fields. Because personal identifiers were unavailable, each notification was counted as one case, and repeated notifications of the same woman could not be identified. Following the clinical classification in the INS~875 form \citep{INS875}, cases were grouped into non-sexual violence, comprising physical violence, psychological violence and neglect (the reference outcome, $k=1$), and sexual violence ($k=2$).

The spatial support was based on Medellín's official cartography of neighbourhoods and rural subdivisions \citep{MedellinCartography}. Of its 332 cartographic units, 47 (20 institutional, 2 unnamed and 25 with a mean female population below 200) were merged into the adjacent unit with which they shared the longest border, leaving $A=285$ analysis units. These comprised 243 urban neighbourhoods in communes 1--16 and 42 units in the five rural districts; for brevity, we refer to all of them as neighbourhoods.

\subsection{Expected counts and observation model}\label{sec:expected}

We used indirect age standardisation to account for differences in neighbourhood age composition and in the age distributions of the two outcomes: sexual violence notifications were concentrated among girls aged 10--14, whereas non-sexual violence notifications were distributed more broadly across adult ages. Age- and sex-structured population denominators were obtained from the official small-area projections of the Medellín Administrative Planning Department \citep{MedellinProjections}.

For each outcome $k$ and five-year age group $a$ (17 groups), we calculated a city-wide reference rate $\omega_{ak}$ by pooling cases and female populations over all neighbourhoods and years. Applying these rates to the age structure of neighbourhood $i=1,\ldots,A$ in year $t=1,\ldots,T$ ($T=5$) gave the expected count
\begin{equation}
E_{ikt} = \sum_{a} N_{iat}\, \omega_{ak}, \qquad
\omega_{ak} = \frac{\sum_{i,t} Y_{iakt}}{\sum_{i,t} N_{iat}},
\label{eq:expected}
\end{equation}
where $Y_{iakt}$ is the observed count for outcome $k$, neighbourhood $i$, age group $a$ and year $t$, and $N_{iat}$ is the corresponding female population. Thus, $E_{ikt}$ is the count expected if that neighbourhood-year experienced the city-wide age-specific notification rates pooled over 2018--2022.

Conditional on the relative risk $r_{ikt}$, the observed counts followed a Poisson model with $\log(E_{ikt})$ as the offset:
\begin{equation}
\begin{gathered}
Y_{ikt} \mid r_{ikt} \sim \text{Poisson}(E_{ikt}\, r_{ikt}), \\[4pt]
\log(E_{ikt}\, r_{ikt}) = \log(E_{ikt}) + \eta_{ikt},
\end{gathered}
\label{eq:pois_st}
\end{equation}
where $\eta_{ikt}=\log(r_{ikt})$ is the log-relative risk for outcome $k$ in neighbourhood $i$ and year $t$, relative to its outcome-specific reference rates.

\subsection{Shared-component model with time-varying spatial loadings}\label{sec:sharedcomp}

We modelled the two log-relative-risk surfaces jointly through a shared-component decomposition \citep{KnorrHeldBest2001,Held2005}. Taking non-sexual violence ($k=1$) as the reference outcome,
\begin{align}
\eta_{i1t} &= \alpha_1 + \delta_{1t}\,\psi_i + \phi_{i1} + \gamma_t + \Delta_{i1t}, \label{eq:eta1}\\
\eta_{i2t} &= \alpha_2 + \delta_{2t}\,\psi_i + \phi_{i2} + \lambda\,\gamma_t + \nu_t + \Delta_{i2t}. \label{eq:eta2}
\end{align}
Each additive term contributes on the log-relative-risk scale, so a value $x$ corresponds to a multiplicative factor $\exp(x)$, holding the remaining terms constant.

The intercepts $\alpha_k$ represent the baseline log-risks. The shared field $\psi_i$ describes a common spatial pattern, assumed constant over 2018--2022, whose contribution to each outcome is scaled by $\delta_{kt}$. Fixing the non-sexual loading at one in 2018--2019 ($\delta_{1,\text{pre}}=1$) anchors this field in units of non-sexual log-risk during that period. The outcome-specific fields $\phi_{ik}$ capture the remaining time-constant spatial structure. A field $\phi_k$ of negligible magnitude therefore indicates that the shared component accounts for most of that outcome's time-constant spatial variation, although interactions may still contribute to its annual risk maps.

The shared temporal effect $\gamma_t$ describes city-wide year-to-year variation, with loading $\lambda$ for sexual violence. The term $\nu_t$ captures additional temporal variation in sexual violence, while $\Delta_{ikt}$ represents neighbourhood-year deviations remaining after the main spatial and temporal effects. We assigned BYM2 structures to the spatial fields \citep{riebler2016bym2}, a second-order random walk (RW2) to $\gamma_t$, and a first-order random walk (RW1) to $\nu_t$. The interaction structures and prior specifications are detailed in Sections~\ref{sec:interactions} and \ref{sec:priors}.

\paragraph{Interpretation of the spatial loadings.}
Following shared-component formulations with outcome- and time-specific loadings \citep{Baker2017}, $\delta_{kt}$ measures how strongly the shared spatial field contributes to outcome $k$ in year $t$. For positive loadings, their ratio $\rho_t=\delta_{2t}/\delta_{1t}$ describes the relative spatial loading: $\rho_t=1$ indicates equal contributions on the log-risk scale, $\rho_t>1$ a stronger contribution to sexual violence, and $0<\rho_t<1$ an attenuated contribution to sexual violence.

A change $D$ in $\delta_{kt}$ multiplies the relative risk in neighbourhood $i$ by $\exp(D\psi_i)$, holding the remaining model terms constant. A positive change raises the contribution to risk where $\psi_i>0$ and lowers it where $\psi_i<0$, steepening the gradient contributed by the shared field. If both positive loadings increase while $\rho_t$ remains stable, this gradient becomes steeper for both outcomes without changing their relative loading. A change in $\rho_t$, by contrast, indicates a change in their relative contribution. These loadings measure amplitude rather than correlation, and their interpretation is conditional on the temporal effects, interactions and prior specification; changes in them do not by themselves imply changes in the correlation between the full risk maps.

Fixing $\delta_{1t}=1$ throughout gives $\rho_t=\delta_{2t}$. Under this restriction, changes in the non-sexual contribution of the shared field cannot be represented through a separate loading and must instead be accommodated by the remaining model components.

\paragraph{Period parameterisation and contrasts.}\label{par:period}
A separate loading for each year and outcome would be informed by only one year of data. We therefore used a reduced-rank parameterisation with $l=3$ distinct loadings per outcome, pooling years within three substantively defined periods:
\begin{equation}
\delta_{kt} =
\begin{cases}
\delta_{k,\text{pre}}  & t \in \{2018,2019\}, \\
\delta_{k,\text{res}} & t \in \{2020,2021\}, \\
\delta_{k,\text{post}} & t = 2022,
\end{cases}
\qquad \delta_{1,\text{pre}}=1.
\label{eq:l3}
\end{equation}
These are the pre-pandemic (2018--2019), restriction (2020--2021) and post-restriction (2022) periods. They are operational labels for the annual data, not measures of individual mobility or restriction intensity, which varied within each period.

The restriction period encompasses Colombia's national mandatory preventive isolation, which began on 25 March 2020 (Decree 457 of 2020), its replacement by selective isolation on 1 September 2020 (Decree 1168 of 2020), and local restrictions continuing into 2021. In-person schooling was suspended from 16 March 2020 \citep{MENSchoolSuspension2020}. Following pilots in 2020, Medellín's public schools resumed alternating attendance on 10 February 2021 \citep{MedellinAlternancia2021}, with a further interruption during April--May 2021 \citep{MedellinReactivacion2021}. Municipal authorities subsequently set 15 July 2021 for the return to full in-person schooling under biosafety protocols \citep{MedellinPresencialidad2021}. By late 2021, Medellín had reopened all economic sectors, and large events had been permitted at full capacity since October \citep{MedellinReopening2021}. For the 2022 school year, municipal authorities announced fully in-person schooling without capacity limits \citep{MedellinPresencialidad2022}. At the annual resolution of our data, this chronology supports distinguishing a pre-pandemic baseline (2018--2019), a period encompassing restrictions and phased reopening (2020--2021), and the first full calendar year following broad reopening (2022).

\paragraph{Period contrasts and posterior inference.}\label{par:contrasts}

The principal contrasts were the restriction-period changes in the two loadings and their ratio:
\begin{equation}
D_k = \delta_{k,\text{res}}-\delta_{k,\text{pre}} \quad (k=1,2), \qquad
D_\rho = \rho_{\text{res}}-\rho_{\text{pre}},
\label{eq:delta_contrast}
\end{equation}
where $D_1=\delta_{1,\text{res}}-1$. For positive loadings, positive $D_1$ and $D_2$ with $D_\rho$ close to zero indicate a steepening common to both outcomes. Secondary contrasts $\delta_{k,\text{post}}-\delta_{k,\text{res}}$ describe the subsequent change, while the final--initial contrasts
\begin{equation}
\delta_{k,\text{post}}-\delta_{k,\text{pre}}, \qquad \rho_{\text{post}}-\rho_{\text{pre}},
\label{eq:delta_post_pre}
\end{equation}
quantify differences from the initial period.

We calculated the loading ratios and period contrasts separately within each of $5{,}000$ draws from the joint posterior distribution of the hyperparameters (seed $511010$), preserving posterior dependence between loadings. We summarised each contrast by its posterior mean, 95\% credible interval and probability of being positive, $\Pr(D>0\mid\mathbf{y})$, where $\mathbf{y}$ denotes the observed counts. A credible interval containing zero was not interpreted as evidence of equivalence between periods, since no equivalence margin had been specified.

To express loading changes on the risk scale, we used Eq.~\eqref{eq:delta_pair} in Appendix~\ref{app:models}. For two neighbourhoods $i$ and $j$, the quantity $\exp[D_k(\psi_i-\psi_j)]$ gives the factor by which the shared component's contribution to their risk ratio changes from the pre-pandemic to the restriction period. When $\psi_i>\psi_j$, a factor above one indicates a widening of this shared spatial contrast, whereas a factor below one indicates a narrowing. Its magnitude depends on both the loading change and the difference between the neighbourhoods' shared-field values; $D_k$ alone therefore cannot be interpreted as a percentage change in risk. We evaluated the risk-scale summaries using joint posterior draws of the loadings and the shared field, propagating uncertainty in both. These summaries describe changes in the contribution of the shared component to between-neighbourhood risk differences, rather than changes in total notification risk.

\subsection{Space-time interactions}\label{sec:interactions}

We considered the four interaction types of \citet{KnorrHeld2000}, which differ in whether variation is structured or unstructured in space and time (Table~\ref{tab:kh}). Structured variation borrows strength from neighbouring areas or adjacent years, whereas unstructured variation imposes no such smoothing. Types~II and IV used an RW2 temporal structure.

To reduce confounding with the main effects, we imposed centring constraints informed by \citet{Goicoa2018}. For Type~I, each outcome's interaction summed to zero and had zero projection on a linear time trend, preventing it from absorbing the city-wide linear component retained by the RW2 main effect. For Type~II, each interaction summed to zero over time within each neighbourhood; the shared RW2 temporal main effect was additionally constrained to have zero projection on a linear time trend. For Type~III, each interaction summed to zero over neighbourhoods within each year. Type~IV combined the neighbourhood-specific temporal sums and year-specific spatial sums. For Types~II and IV, an additive diagonal term $10^{-4}\mathbf{I}_{AT}$ regularised the intrinsic directions remaining after these constraints.

Type~I assigns a Gaussian effect to each neighbourhood-year-outcome observation, accounting for observation-level overdispersion. Integrating out these effects yields a Poisson-lognormal mixture, while the conditional observation model remains Poisson. Its scale $\sigma$ refers to the Gaussian prior before conditioning on the constraints and thus approximates, rather than equals, the marginal standard deviation of the constrained effects. The primary M4 specification used a Type~I interaction. The comparison of interaction types and the numerical issues affecting Types~II and IV are reported in Section~\ref{sec:modelcomparison}.

\begin{table*}[pos=h]
\centering
\small
\caption{The four space--time interaction types, by their spatial and temporal structure.}
\label{tab:kh}
\begin{tabular*}{\tblwidth}{@{}LLLL@{}}
\toprule
Type & Space & Time & Captures \\
\midrule
I   & unstructured      & unstructured & independent area--year deviations (overdispersion) \\
II  & unstructured      & RW2           & area-specific temporal trends \\
III & structured (iCAR) & unstructured & year-specific spatial patterns \\
IV  & structured (iCAR) & RW2           & smoothly evolving spatial patterns \\
\bottomrule
\end{tabular*}
\end{table*}

\subsection{Prior specification}\label{sec:priors}

The BYM2 specification \citep{riebler2016bym2} represents each shared or outcome-specific spatial field as $\theta=\sigma(\sqrt{\pi}\,\mathbf{u}+\sqrt{1-\pi}\,\mathbf{v})$, where $\mathbf{u}$ is a scaled intrinsic CAR component on the neighbourhood adjacency graph and $\mathbf{v}$ is unstructured heterogeneity. The parameter $\sigma$ controls the overall scale, while $\pi\in(0,1)$ controls the balance between structured and unstructured variation. These parameters describe the magnitude and spatial structure of variation, rather than the locations of high and low values: $\sigma=0.5$, for example, corresponds to a typical multiplicative contribution of about $\exp(0.5)\approx1.6$, while $\pi$ near one favours a smooth, clustered pattern and $\pi$ near zero favours unstructured neighbourhood-to-neighbourhood variation.

Table~\ref{tab:priors} in Appendix~\ref{app:models} summarises the priors. In the BYM2 shared-component models, penalised-complexity (PC) priors \citep{simpson2017pc} shrink the latent components towards simpler base models unless the data support greater complexity. We specified $P(\sigma>1)=0.01$ for the standard deviations of the BYM2 fields, temporal random walks and space-time interactions. For the BYM2 mixing parameter, we set $P(\pi<0.5)=2/3$, mildly favouring unstructured over structured variation. The free spatial loadings $\delta_{kt}$ and the temporal loading $\lambda$ received weakly informative $\mathcal{N}(0,1)$ priors. The interaction loading $\varrho_t$, used only in the alternative shared-interaction model, received a $\text{Gamma}(10,10)$ prior (shape and rate), centred at one with a 95\% prior interval of $0.48$--$1.71$.

All intrinsic components were constrained to sum to zero. Both temporal main effects were scaled, and the RW2 precision included an additive diagonal term $10^{-4}\mathbf{I}_T$, retaining a regularised linear component except where the additional Type~II constraint applied.

\subsection{Estimation and model comparison}\label{sec:estimation}

All models were fitted using integrated nested Laplace approximation (INLA) \citep{Rue2009} in the \texttt{R-INLA} package \citep{INLApackage}. In the main specification, a common spatial field contributed to both outcomes through period-specific loadings. An alternative formulation, motivated by \citet{Retegui2024}, allowed the relative contribution of a shared space-time interaction to vary between outcomes over time.

We compared alternative specifications using the deviance information criterion (DIC) \citep{spiegelhalter2002} and the Watanabe--Akaike information criterion (WAIC) \citep{watanabe2010}. Lower values indicate a preferable balance between fit and complexity under each criterion. Model formulations and criterion definitions are provided in Appendix~\ref{app:models}.

For each neighbourhood-year, we exponentiated the posterior mean log-relative risk to obtain a point estimate of $r_{ikt}=\exp(\eta_{ikt})$, and reported a 95\% credible interval. We also identified neighbourhoods where both relative risks exceeded one relative to their respective pooled references. Using $1{,}000$ joint posterior samples, we estimated
\begin{equation}
p_{it} = \Pr\!\left(r_{i1t}>1 \ \text{and}\ r_{i2t}>1 \mid \mathbf{y}\right),
\label{eq:exceed}
\end{equation}
accounting for posterior dependence between the two risk surfaces. Annual dual hotspots were defined by $p_{it}>0.8$, while persistent dual hotspots required $p_{it}>0.95$ in every year. The latter criterion applies separately to each year and does not imply a posterior probability above $0.95$ of simultaneous exceedance across all five years.

We assessed numerical stability by repeating estimation from different initial values and examined sensitivity to the loading and spatial priors. We also varied the scale of the sexual-specific spatial field or omitted it to assess how the separation of shared and outcome-specific variation affected inference. These analyses focused on the stability of the period contrasts. Detailed settings are provided in Appendix~\ref{app:models} (Tables~\ref{tab:appA_starts} and~\ref{tab:appA_m4_sens}).

To assess sensitivity to the composition and source of notifications, we repeated the main analysis restricting cases and population denominators to women aged 20 years and over, excluding neglect and abandonment from non-sexual violence, and using only notifications from health institutions. The age threshold was the youngest boundary of the five-year population bands that excluded all minors. To identify the notifying institution, we linked the cleaned records to the original MEData export using notification date, age, type of violence and other unchanged fields; $99\%$ of records were linked. Expected counts were recalculated for each analysis.

Exploratory stratified analyses considered intrafamilial notifications involving a partner, ex-partner or relative ($32{,}152$, $69.3\%$) and the remaining notifications ($14{,}238$, $30.7\%$). The latter included acquaintances, non-cohabiting partners, strangers, and unknown or missing relationships. We fitted the main specification to each stratum using all notifications and, separately, notifications from health institutions only. Expected counts were calculated from stratum- and outcome-specific crude rates, without age standardisation. Each analysis estimated its own shared field, so the loadings describe the spatial pattern within that stratum and should not be compared directly across strata.

\subsection{Model checking}\label{sec:modelchecking}

We assessed the fit and predictive performance of the main specification using three complementary checks. For each outcome, we examined the leave-one-out predictive distributions computed by INLA using the probability integral transform (PIT). To account for the discrete nature of the counts, we constructed the nonrandomized PIT histogram of \citet{czado2009}, which distributes each observation's contribution uniformly over the interval between $\Pr(Y\le y-1)$ and $\Pr(Y\le y)$.

We also generated 500 datasets from the posterior predictive distribution and compared their count distributions with those observed, assessing whether the model reproduced the distributional features of each outcome.

For an internal predictive validation, we randomly withheld 15\% of the $2{,}850$ neighbourhood-year-outcome cells (428 cells). We recalculated reference rates and expected counts using only the remaining cells and refitted the model. We then assessed the proportion of withheld counts falling within their 95\% posterior predictive intervals.

\subsection{Ethical considerations}\label{sec:ethics}

This study used secondary surveillance data obtained from MEData, Medell\'in's municipal open-data platform. The records were already anonymised when accessed by the researchers, and no direct personal identifiers were available. The study involved no contact with the individuals represented in the records and no intervention. For statistical modelling, notifications were aggregated by neighbourhood, year and type of violence, and findings are reported as aggregate summaries and area-level estimates. Given the sensitive nature of the information, case-level notification records are not redistributed with the analysis code. The reported spatial patterns describe notifications at the neighbourhood level and should not be interpreted as individual risk estimates or as characteristics of residents.

\subsection{Descriptive patterns}\label{sec:descriptive}

\begin{figure}[pos=htbp]
    \centering
    \includegraphics[width=0.95\linewidth]{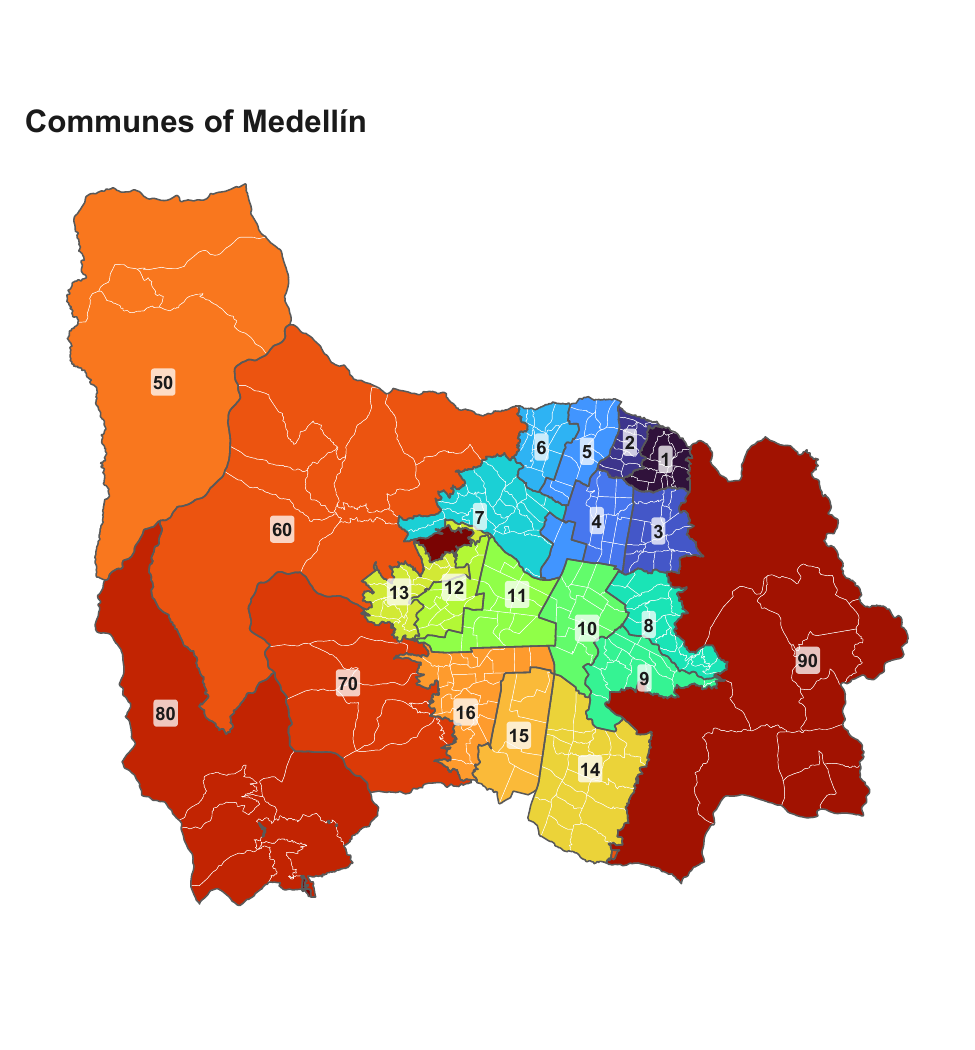}
    \caption{Administrative division of Medellín into communes and rural districts, which contain the 285 analysis units.}
    \label{fig:region}
\end{figure}

Between 2018 and 2022, $46{,}390$ suspected cases of violence against women were geocoded to the $285$ analysis units of Medellín: $34{,}072$ ($73\%$) were classified as non-sexual and $12{,}318$ ($27\%$) as sexual. Fig~\ref{fig:region} and Table~\ref{tab:communes} show the communes and rural districts into which these units are grouped.

\begin{table}[pos=h]
\centering
\small
\caption{Communes and rural districts of Medellín, keyed to the numbers in Fig~\ref{fig:region}.}
\label{tab:communes}
\begin{tabular*}{\tblwidth}{@{}LL@{\qquad}LL@{}}
\toprule
\# & Commune & \# & Commune \\
\midrule
1  & Popular            & 12 & La América \\
2  & Santa Cruz         & 13 & San Javier \\
3  & Manrique           & 14 & El Poblado \\
4  & Aranjuez           & 15 & Guayabal \\
5  & Castilla           & 16 & Belén \\
6  & Doce de Octubre    & 50 & San Sebastián de Palmitas \\
7  & Robledo            & 60 & San Cristóbal \\
8  & Villa Hermosa      & 70 & Altavista \\
9  & Buenos Aires       & 80 & San Antonio de Prado \\
10 & La Candelaria      & 90 & Santa Elena \\
11 & Laureles--Estadio  &    & \\
\bottomrule
\end{tabular*}
\end{table}

The outcomes differed in their age profiles. Non-sexual violence notifications were distributed across adult ages, whereas sexual violence notifications were concentrated among girls and adolescents, peaking at ages 10--14. The age composition of non-sexual notifications also changed over time: the proportion involving girls under 18 declined from $32.6\%$ ($6{,}664/20{,}454$) in 2018--2020 to $7.8\%$ ($1{,}067/13{,}618$) in 2021--2022.

This shift was concentrated in notifications from the municipal family commissaries (Comisarías de Familia), which accounted for about four in five non-sexual notifications involving girls in 2018--2020 and almost no sexual violence notifications. The commissaries reported no cases from August to December 2020. From January 2021, notifications involving adult women resumed, but those involving girls remained uncommon: $71$ in 2021 and $80$ in 2022, compared with $2{,}307$ in 2019. Among health institutions, the proportion involving girls was similar in the two periods ($19.3\%$ and $19.6\%$). The change in age composition was therefore associated with a change in commissary notifications, whose administrative cause is undocumented.

\begin{figure*}[pos=tp]
    \centering
    \includegraphics[width=0.85\textwidth]{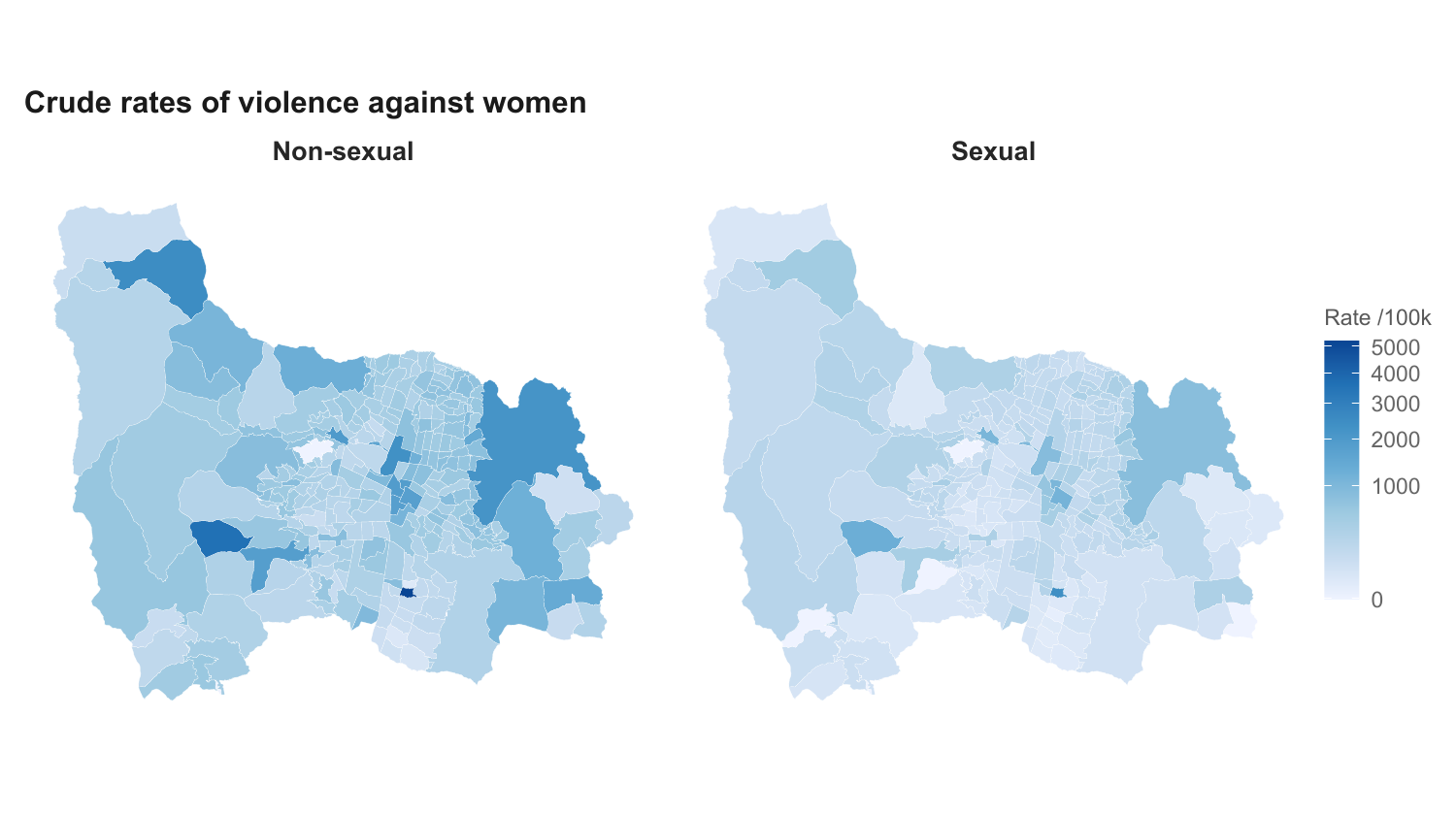}
    \caption{Crude notification rates per 100{,}000 women, pooled over 2018--2022, by type of violence.}
    \label{fig:crude}
\end{figure*}

Crude rates per 100{,}000 women, calculated as total notifications divided by total person-years over 2018--2022, showed marked spatial heterogeneity (Fig~\ref{fig:crude}). Both outcomes followed broadly similar patterns, with the highest rates in the central commune of La Candelaria and the north-eastern communes of Popular, Santa Cruz and Manrique. The more affluent south-eastern commune of El Poblado was among those with the lowest rates.

\begin{figure}[pos=htbp]
    \centering
    \includegraphics[width=\linewidth]{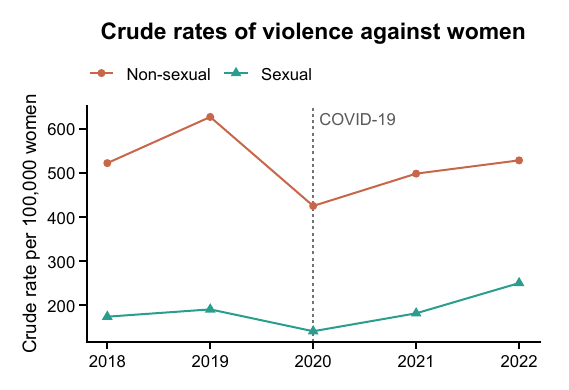}
    \caption{City-level crude notification rates per 100{,}000 women, by year and type of violence.}
    \label{fig:crude_trend}
\end{figure}

\begin{figure}[pos=htbp]
    \centering
    \includegraphics[width=\linewidth]{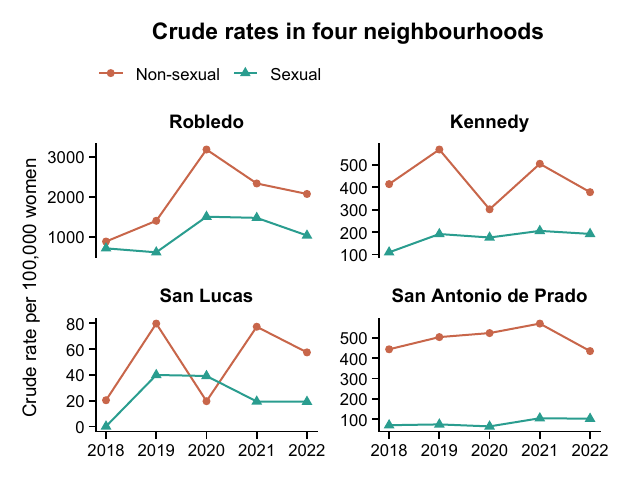}
    \caption{Crude notification rates in four selected neighbourhoods, by year and type of violence.}
    \label{fig:crude_areas}
\end{figure}

At the city level, both rates declined in 2020 and increased thereafter (Fig~\ref{fig:crude_trend}). The initial decline was sharper for non-sexual violence, while sexual violence reached its highest rate in the series in 2022. Temporal patterns nevertheless varied substantially across neighbourhoods (Fig~\ref{fig:crude_areas}). Because crude rates are unstable in small populations, they are used here descriptively; the main model-based analysis uses smoothed, age-standardised relative risks.

\subsection{Model comparison}\label{sec:modelcomparison}

We compared the main specification, in which both spatial loadings varied by period (M4), with alternative representations of the joint spatial and temporal variation (Appendix~\ref{app:models}). These included a separable baseline with separate between-outcome correlations for spatial fields and temporal trends (M0), a shared-component model with a constant spatial loading (M1), and extensions incorporating the four Knorr-Held interaction structures (M2-I to M2-IV). We also considered specifications in which only the sexual-violence loading varied, either annually or across the three periods, while the non-sexual loading remained fixed at one (M3). A further alternative, motivated by \citet{Retegui2024}, allowed the loading of a shared space-time interaction to vary over time while keeping the spatial loading constant (SCM).

Adding a Type~I interaction to the shared-component model with a constant spatial loading reduced WAIC from $16{,}305$ to $15{,}268$ (Table~\ref{tab:models}). The Type~II and IV specifications produced unresolved numerical warnings and were excluded from primary inference. Among the retained specifications, WAIC was lowest when both spatial loadings varied by period ($15{,}253$), followed by models in which only the sexual-violence loading varied annually ($15{,}258$) or by period ($15{,}260$). DIC was lowest for the specification combining a constant spatial loading with a Type~III interaction.

\begin{table*}[pos=h]
\centering
\small
\caption{Model comparison on the same $2{,}850$ neighbourhood--year--outcome cells, with fixed expected counts. Lower values are preferred under each criterion. $p_{\mathrm{WAIC}}$ is the WAIC effective number of parameters. Bold identifies the lowest WAIC and DIC; $^{\dagger}$indicates the model used for the main analysis.}
\label{tab:models}
\begin{tabular*}{\tblwidth}{@{}LLLL@{}}
\toprule
Model & WAIC & DIC & $p_{\mathrm{WAIC}}$ \\
\midrule
M0: separable baseline & 16{,}314 & 15{,}981 & 672 \\
M1: static shared components & 16{,}305 & 15{,}976 & 664 \\
M2-I: centred Type I & 15{,}268 & 15{,}271 & 881 \\
M2-III: centred Type III & 15{,}304 & \textbf{15{,}243} & 851 \\
M3: annual $\delta_t$ ($l=T$) & 15{,}258 & 15{,}262 & 873 \\
M3: three-period $\delta_t$ ($l=3$) & 15{,}260 & 15{,}263 & 874 \\
M4: three-period loadings in both outcomes$^{\dagger}$ & \textbf{15{,}253} & 15{,}251 & 867 \\
SCM: shared interaction $\varrho_t$ & 15{,}373 & 15{,}273 & 836 \\
\bottomrule
\end{tabular*}
\end{table*}

We assessed uncertainty in WAIC differences using standard errors calculated from the paired pointwise contributions \citep{vehtari2017}. Relative to M4, WAIC was $1{,}052$ units higher for M1 and $1{,}060$ higher for M0 (standard errors $101$), and $119$ higher for SCM (standard error $37$). Differences with M2-I ($+15$, standard error $15$), M2-III ($+51$, standard error $32$), annual-loading M3 ($+5$, standard error $13$) and three-period M3 ($+7$, standard error $12$) were within two standard errors. We prioritised WAIC over DIC because it integrates the predictive density over the posterior rather than evaluating fit at a point estimate \citep{watanabe2010,gelman2014}; the lower DIC of M2-III was therefore not decisive.

WAIC did not clearly distinguish M4 from M2-I, M2-III or either version of M3. Among these models, M4 directly represents period changes in both spatial loadings, allowing a common steepening to be distinguished from a change in their relative contribution. We therefore selected M4 for the main analysis. Annual and three-period M3 also differed little in WAIC ($15{,}258$ versus $15{,}260$; standard error of the difference $5$), supporting the more parsimonious period-based representation. Unless otherwise specified, the following results refer to three-period M4.

Both annual and three-period M3 produced fitted risk surfaces very similar to those of M4. Correlations between their posterior mean log-relative risks and those of M4 were $0.997$ for non-sexual violence under both specifications, and $0.9992$ and $0.9996$ for sexual violence under annual and three-period M3, respectively. The correlation between the joint exceedance probabilities of three-period M3 and M4 was $0.999$. Despite this close agreement in fitted risks, the models allow different decompositions of temporal changes, as examined below. M4 and both versions of M3 had lower WAIC and DIC than SCM, although differences in several model components prevent attributing this result to any single component.

Recorded fitting times are reported in Table~\ref{tab:cpu} (Appendix~\ref{app:models}).

\subsection{Model adequacy}\label{sec:adequacy}

\begin{figure}[pos=htbp]
    \centering
    \includegraphics[width=\linewidth]{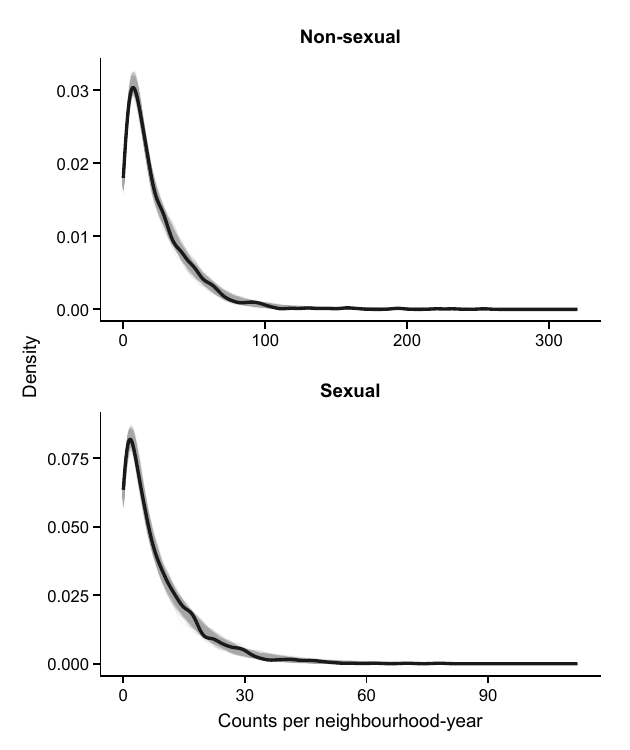}
    \caption{Posterior predictive distributions of counts under M4. The observed count density (dark line) is compared with 500 replicated datasets (light grey), separately for each outcome.}
    \label{fig:ppc}
\end{figure}

The 500 posterior predictive datasets reproduced the main features of the observed count distributions, including totals, maxima and 99th percentiles (Fig~\ref{fig:ppc}). The proportion of zero counts for sexual violence was also closely reproduced (11.3\% observed versus 10.5\% on average in the replicated datasets). For non-sexual violence, the observed proportion of zeros was higher than predicted: 48 of $1{,}425$ neighbourhood-years (3.4\%) had no notifications, compared with a replicated mean of 2.3\% (95\% interval 1.5--3.1\%). These zeros occurred in every year (8--11 annually), with 35 of the 48 in rural areas and the remainder in neighbourhoods with small expected counts.

\begin{figure}[pos=htbp]
    \centering
    \includegraphics[width=\linewidth]{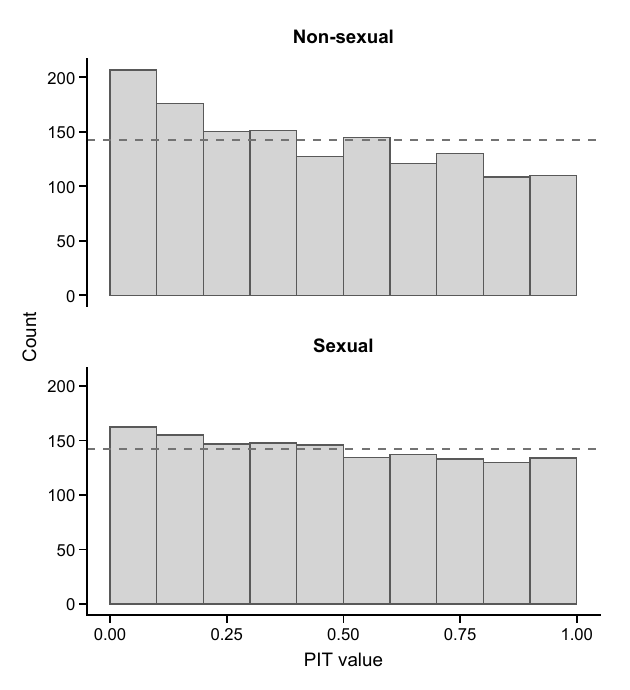}
    \caption{Nonrandomized PIT histograms \citep{czado2009} of the leave-one-out predictive distributions under M4, separately for each outcome. The dashed line indicates the expected frequency under uniformity.}
    \label{fig:pit}
\end{figure}

Leave-one-out predictive calibration differed between outcomes (Fig~\ref{fig:pit}). For sexual violence, the nonrandomized PIT histogram was close to uniform, with 22\% of its mass below 0.2 and 19\% above 0.8, compared with 20\% expected in each tail. For non-sexual violence, the corresponding proportions were 27\% and 15\%, indicating a tendency for observed counts to fall below the centre of their predictive distributions. The excess of low PIT values for non-sexual violence persisted among positive counts, with 25\% of the mass below 0.2. Mean mid-point PIT values were similar across periods ($0.45$, $0.45$ and $0.43$) and tertiles of the shared field ($0.44$--$0.45$), compared with an expectation of 0.5 under calibration when all counts are included. The effect of this calibration departure on the estimated loading contrasts was not directly quantified.

\begin{figure}[pos=htbp]
    \centering
    \includegraphics[width=\linewidth]{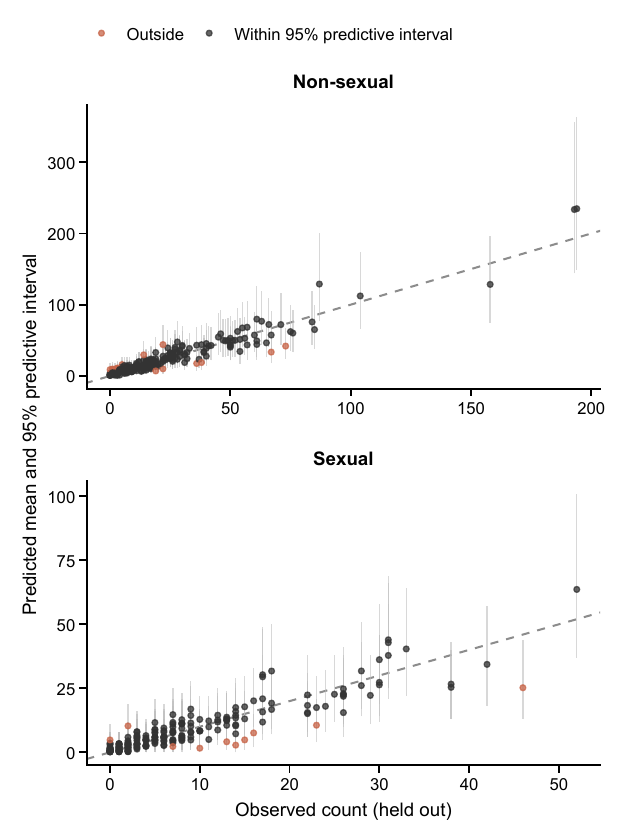}
    \caption{Internal hold-out validation of M4. Observed and predicted counts for the 15\% of neighbourhood--year--outcome cells withheld from fitting, separately for each outcome. The dashed line marks equality, and colours indicate whether the 95\% posterior predictive interval contains the observed count.}
    \label{fig:holdout}
\end{figure}

In the internal hold-out validation, predictive interval coverage was close to the nominal 95\% level. The intervals covered 403 of the 428 withheld counts (94\%), with coverage of 93\% for non-sexual and 95\% for sexual violence (Fig~\ref{fig:holdout}).

\subsection{Time-varying spatial loadings}\label{sec:timevarying}

\begin{figure}[pos=htbp]
    \centering
    \includegraphics[width=\linewidth]{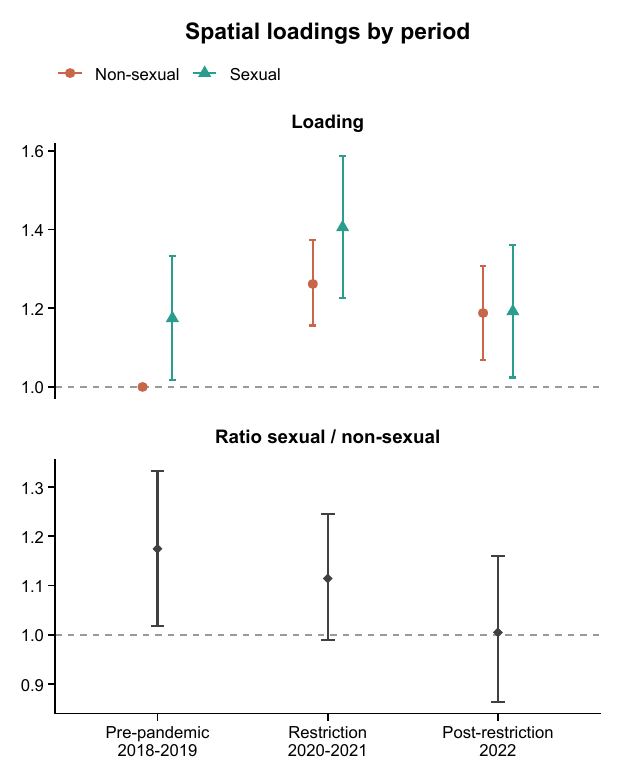}
    \caption{Spatial loadings of the shared field by period in M4, for non-sexual ($\delta_{1t}$) and sexual violence ($\delta_{2t}$), and their ratio $\rho_t$, with 95\% credible intervals from $5{,}000$ joint hyperparameter draws. The non-sexual loading is fixed at one in 2018--2019; $\rho_t=1$ indicates equal contributions of the shared component on the log-risk scale.}
    \label{fig:coupling_periods}
\end{figure}

During the 2020--2021 restriction period, the spatial loading increased for both outcomes (Fig~\ref{fig:coupling_periods}, Table~\ref{tab:period_contrasts}). The increase was $0.26$ (95\% CrI $0.16$--$0.37$) for non-sexual violence and $0.23$ ($0.09$--$0.37$) for sexual violence, both with posterior probabilities above $0.999$. Their ratio showed no clear change ($-0.06$, $-0.21$ to $0.10$). These estimates indicate a strengthening of the shared spatial contribution to both outcomes, without clear evidence of a change in their relative loading.

Following the restriction period, the sexual-violence loading decreased by $0.21$ ($0.06$--$0.37$), leaving its 2022 estimate close to the pre-pandemic value (difference $0.02$, $-0.13$ to $0.17$). The non-sexual loading remained elevated, at $0.19$ ($0.07$--$0.31$) above its initial value. The loading ratio was estimated to be lower in 2022 than before the pandemic, although its contrast interval included zero ($-0.17$, $-0.34$ to $0.01$).

\begin{table*}[pos=h]
\centering
\small
\caption{Spatial loadings by period in M4, their ratio $\rho_t=\delta_{2t}/\delta_{1t}$, and period contrasts. The non-sexual loading is fixed at one in 2018--2019, setting the scale of the shared field. Entries are posterior means and 95\% credible intervals from $5{,}000$ joint hyperparameter draws, with each ratio and difference evaluated within a draw; $\Pr(>0)$ is the posterior probability that the contrast is positive.}
\label{tab:period_contrasts}
\begin{tabular*}{\tblwidth}{@{}LLLL@{}}
\toprule
 & Non-sexual, $\delta_{1t}$ & Sexual, $\delta_{2t}$ & Ratio, $\rho_t$ \\
\midrule
2018--2019 & 1 (fixed) & 1.17 (1.02, 1.33) & 1.17 (1.02, 1.33) \\
2020--2021 & 1.26 (1.16, 1.37) & 1.41 (1.23, 1.59) & 1.11 (0.99, 1.24) \\
2022 & 1.19 (1.07, 1.31) & 1.19 (1.02, 1.36) & 1.00 (0.86, 1.16) \\
\addlinespace
Restriction $-$ pre & $+0.26$ (0.16, 0.37) & $+0.23$ (0.09, 0.37) & $-0.06$ ($-0.21$, 0.10) \\
\quad $\Pr(>0)$ & $>0.999$ & $>0.999$ & 0.22 \\
Post $-$ restriction & $-0.07$ ($-0.19$, 0.04) & $-0.21$ ($-0.37$, $-0.06$) & $-0.11$ ($-0.26$, 0.05) \\
\quad $\Pr(>0)$ & 0.11 & 0.004 & 0.08 \\
Post $-$ pre & $+0.19$ (0.07, 0.31) & $+0.02$ ($-0.13$, 0.17) & $-0.17$ ($-0.34$, 0.01) \\
\quad $\Pr(>0)$ & 0.999 & 0.59 & 0.03 \\
\bottomrule
\end{tabular*}
\end{table*}

On the risk scale, the shared component's contribution to the risk ratio between the 90th and 10th percentiles of the shared field widened during restrictions by about $36\%$ for non-sexual violence (factor $1.36$, 95\% CrI $1.21$--$1.51$) and $31\%$ for sexual violence ($1.31$, $1.11$--$1.54$). In 2022, this contribution remained about $25\%$ wider than before the pandemic for non-sexual violence ($1.25$, $1.09$--$1.42$), whereas the estimate for sexual violence was close to its initial value ($1.02$, $0.86$--$1.21$). These summaries propagated uncertainty in both the loadings and the shared field through joint posterior draws (Appendix~\ref{app:psi_joint}). The percentiles were recalculated within each draw and therefore represent positions in the field rather than a fixed pair of neighbourhoods. The percentages refer specifically to changes in the shared component's contribution to the risk ratio.

\paragraph{Sensitivity analyses.}

The period contrasts were stable across alternative starting values, fixed values for the scale of the sexual-specific field, and replacement of the Type~I interaction with Type~III (Tables~\ref{tab:appA_starts} and~\ref{tab:appA_m4_sens}). Absolute levels of the sexual loading and its ratio varied more than the period contrasts. The comparison with M3 also showed how the loading constraints affected interpretation: with the non-sexual loading fixed at one, three-period M3 estimated an increase in the relative loading during restrictions (Table~\ref{tab:appA_m3_periods}), whereas M4 represented the change through increases in both loadings.

Restricting the analysis to notifications from health institutions preserved the main pattern. The restriction-period increases were $0.30$ (95\% CrI $0.18$--$0.41$) for non-sexual violence and $0.22$ ($0.09$--$0.35$) for sexual violence, with no clear change in their ratio ($-0.08$, $-0.21$ to $0.06$). The non-sexual loading remained elevated in 2022, while the sexual loading was estimated close to its initial value. Similar patterns were obtained among women aged 20 years and over and after excluding neglect and abandonment from non-sexual violence (Table~\ref{tab:appA_m4_sens}).

Exploratory analyses by perpetrator relationship suggested opposite changes in the loading ratio for intrafamilial and other notifications. However, neither stratum showed clear evidence of a ratio change when restricted to health-institution notifications (Table~\ref{tab:appA_m4_sens}). These differences thus depended on the reporting source; the stratified analyses also used crude reference rates and separately estimated shared fields.

\subsection{Spatial and temporal components}\label{sec:posterior}

The shared spatial field combined spatially structured and unstructured variation and had a larger estimated scale than either outcome-specific field. Full posterior summaries are reported in Table~\ref{tab:appA_hyper_full} (Appendix~\ref{app:models}). These scales describe variability in the latent fields; their contribution to each outcome also depends on the corresponding loading.

For sexual violence, the absolute loading levels were sensitive to the allocation of spatial variation between the shared and outcome-specific fields. Refits with different fixed values for the sexual-specific scale produced similar fits and stable period contrasts (Table~\ref{tab:appA_starts}).

\begin{figure}[pos=htbp]
    \centering
    \includegraphics[width=0.75\linewidth]{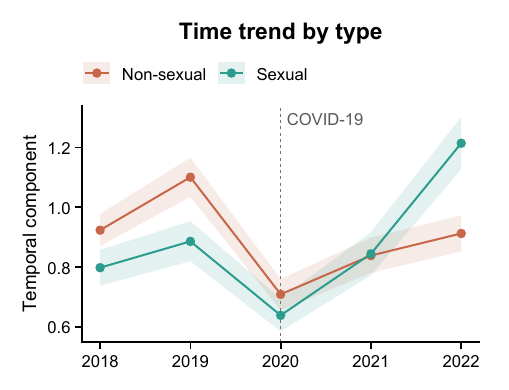}
    \caption{Multiplicative factors combining the intercept and temporal components, with 95\% credible intervals: $\exp(\alpha_1+\gamma_t)$ for non-sexual violence and $\exp(\alpha_2+\lambda\gamma_t+\nu_t)$ for sexual violence. Spatial effects and space-time interactions are excluded.}
    \label{fig:temporal}
\end{figure}

To summarise the temporal pattern, we combined the intercept and temporal effects for each outcome (Fig~\ref{fig:temporal}). Both resulting multiplicative factors reached their minimum in 2020: $0.71$ (95\% CrI $0.66$--$0.76$) for non-sexual violence and $0.64$ ($0.59$--$0.70$) for sexual violence. By 2022, they had increased to $0.91$ ($0.85$--$0.97$) and $1.21$ ($1.13$--$1.30$), respectively. These factors describe the temporal component common to neighbourhoods within each outcome; full neighbourhood risks also incorporate spatial effects and space-time interactions.

The space-time interactions represented local departures from the spatial patterns and temporal trends, with similar estimated scales for the two outcomes. These terms allowed for extra-Poisson variation, consistent with the improvement in fit after adding a Type~I interaction to the model with constant spatial loadings (Table~\ref{tab:models}).

\subsection{Relative-risk surfaces and their decomposition}\label{sec:decomposition}

\begin{figure*}[pos=tp]
    \centering
    \includegraphics[width=0.6\textwidth]{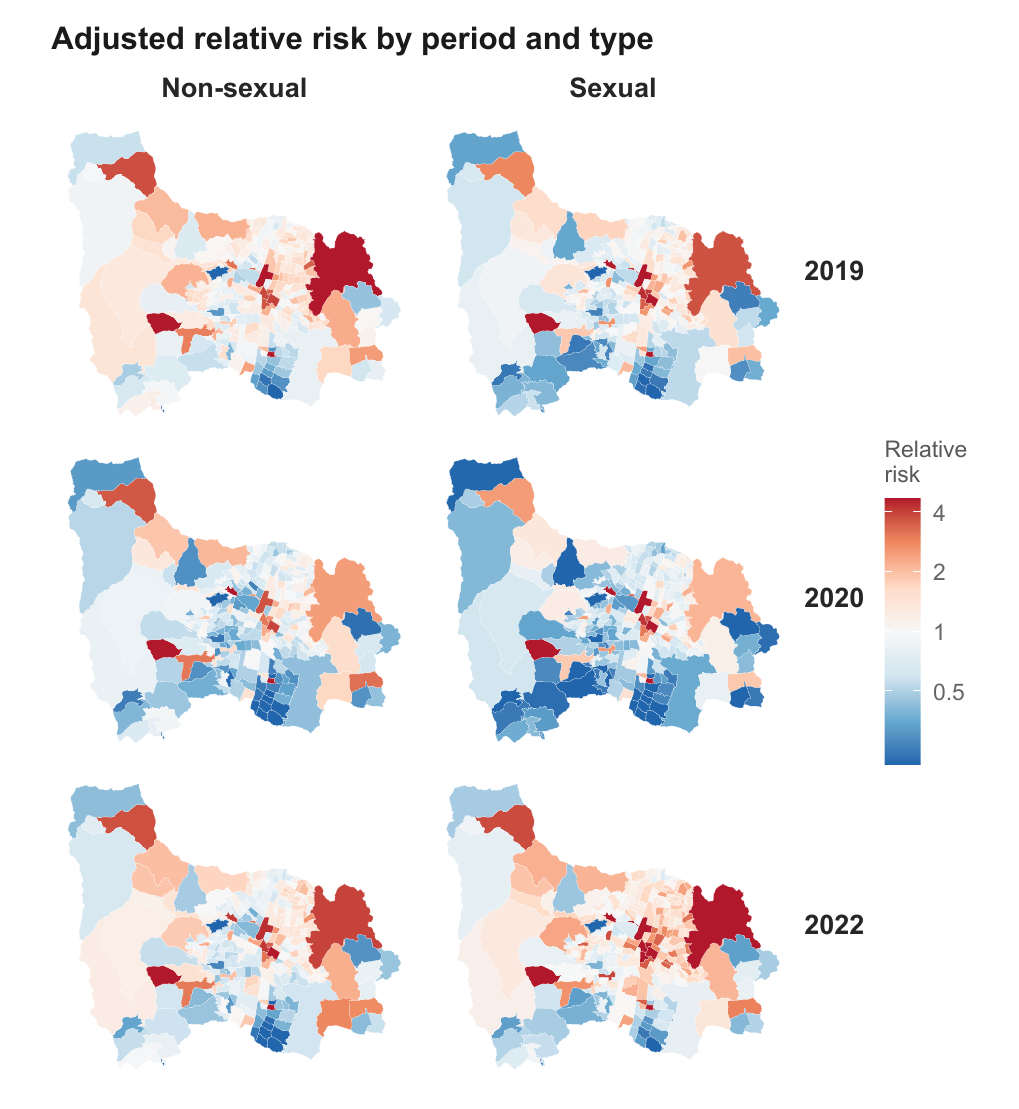}
    \caption{Estimated relative risks by neighbourhood and outcome in M4, for 2019, 2020 and 2022. Point estimates exponentiate posterior mean log-relative risks and incorporate all model components. Red indicates risk above one and blue below one, relative to expected counts based on outcome-specific, age-specific rates pooled over 2018--2022. Colours use a log scale.}
    \label{fig:rr}
\end{figure*}

\begin{figure}[pos=H]
    \centering
    \includegraphics[width=1.1\linewidth]{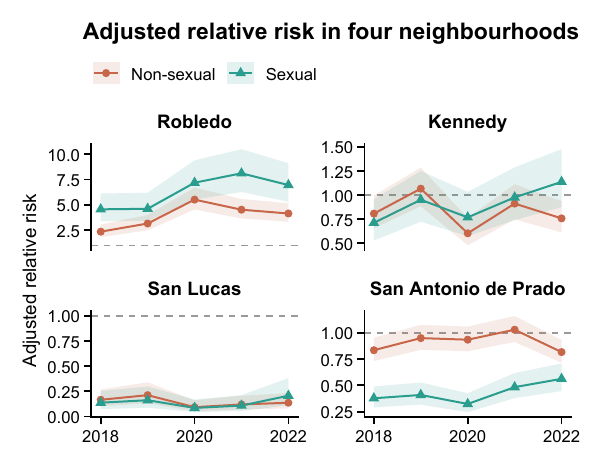}
    \caption{Estimated relative risks and 95\% credible intervals in the four neighbourhoods of Fig~\ref{fig:crude_areas}.}
    \label{fig:areas_fitted}
\end{figure}

Fig~\ref{fig:rr} shows the full relative-risk surfaces for one year in each period (2019, 2020 and 2022), incorporating all model components. The complete five-year series is provided in Fig~\ref{fig:appA_rr_full} (Appendix~\ref{app:models}). In 2020, estimated sexual-violence risk was below the pooled reference across much of the west and south-west, partly reflecting the lower city-wide temporal component that year. Elevated risk nevertheless persisted in several neighbourhoods with high non-sexual risk. The trajectories in the four neighbourhoods selected to illustrate crude rates show how the estimated risks of both outcomes evolved locally (Fig~\ref{fig:areas_fitted}).

\begin{figure*}[pos=tp]
    \centering
    \includegraphics[width=0.75\textwidth]{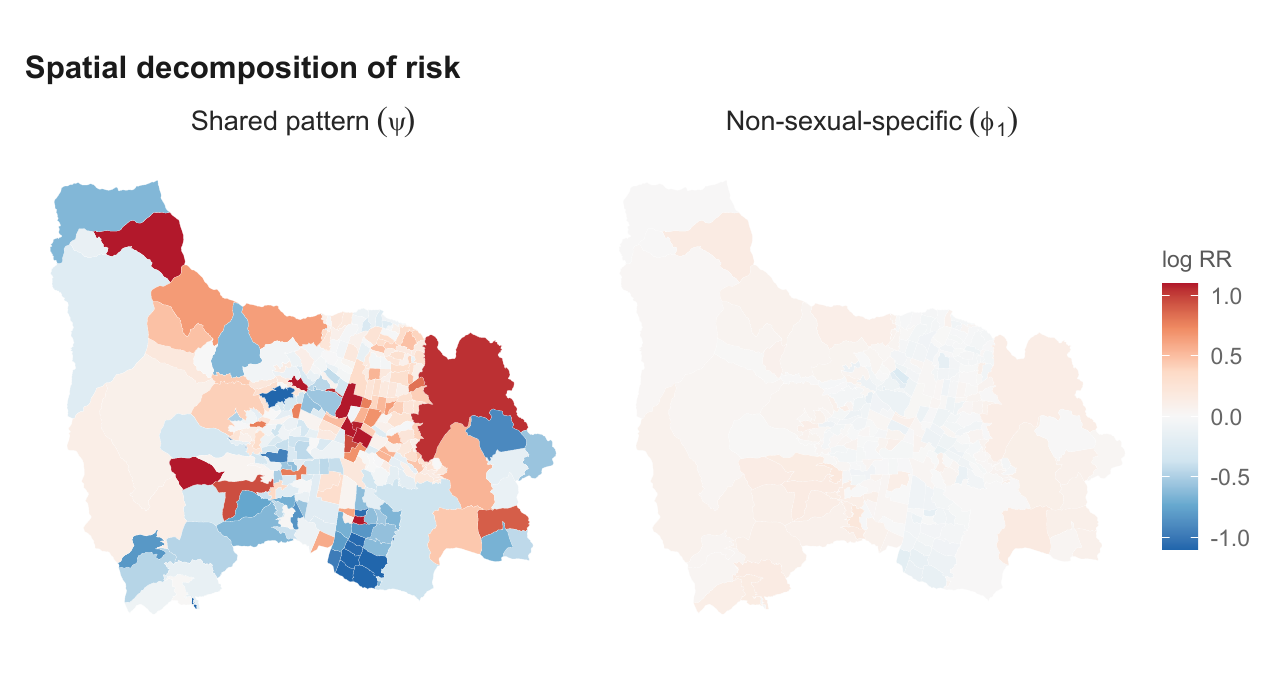}
    \caption{Posterior mean spatial fields in M4: the shared field $\psi$ and the non-sexual-specific field $\phi_1$. Values are on the log-risk scale. The shared contribution is $\delta_{kt}\psi_i$ for outcome $k$ in period $t$, with $\delta_{1,\text{pre}}=1$; $\phi_1$ enters the non-sexual predictor directly.}
    \label{fig:components}
\end{figure*}

The shared field describes the spatial pattern underlying both outcomes (Fig~\ref{fig:components}). With $\delta_{1,\text{pre}}=1$, its multiplicative contribution to non-sexual risk in 2018--2019 is $\exp(\psi_i)$. This contribution was typically highest in La Candelaria (about $1.5$), followed by Manrique, Aranjuez, Popular and Villa Hermosa ($1.2$--$1.4$). The lowest values were in El Poblado ($0.56$), San Antonio de Prado and Bel\'en (both $0.77$). These values describe the shared contribution rather than the full relative risks shown above.

Higher shared-field values were associated with poorer living conditions across the 16 urban communes. In a post hoc comparison of commune averages of $\psi$ with the 2018 Multidimensional Living Conditions Index (IMCV) \citep{MedellinIMCV}, the rank correlation was $-0.65$ (95\% CrI $-0.74$ to $-0.52$), calculated using $2{,}000$ joint posterior draws. La Candelaria was the main exception, combining a relatively high index with the highest shared-field values. Rural districts combined low index values with shared-field values at or below the city average. This association is descriptive and was not included in the model.

The non-sexual-specific field added a smaller spatial pattern (Fig~\ref{fig:components}). Its mean was slightly positive across the 42 rural units ($0.08$), compared with $-0.01$ across the 243 urban neighbourhoods, indicating a modest additional contribution in rural areas. The sexual-specific field is not presented separately because its allocation relative to the shared field was less clearly identified (Section~\ref{sec:posterior}).

\subsection{Dual hotspots}\label{sec:hotspots}

For each neighbourhood $i$ and year $t$, we calculated the posterior probability that both notification relative risks exceeded one:
\[
p_{it}
=
\Pr\!\left(
r_{i1t}>1,\ r_{i2t}>1
\mid \mathbf{y}
\right),
\]
where $r_{i1t}$ and $r_{i2t}$ are the full relative risks for non-sexual and sexual violence, respectively, and $\mathbf{y}$ denotes the observed counts. Exceedance is defined relative to the outcome-specific, age-specific reference rates pooled over 2018--2022.

\begin{figure*}[pos=htbp]
    \centering
    \includegraphics[width=0.75\textwidth]{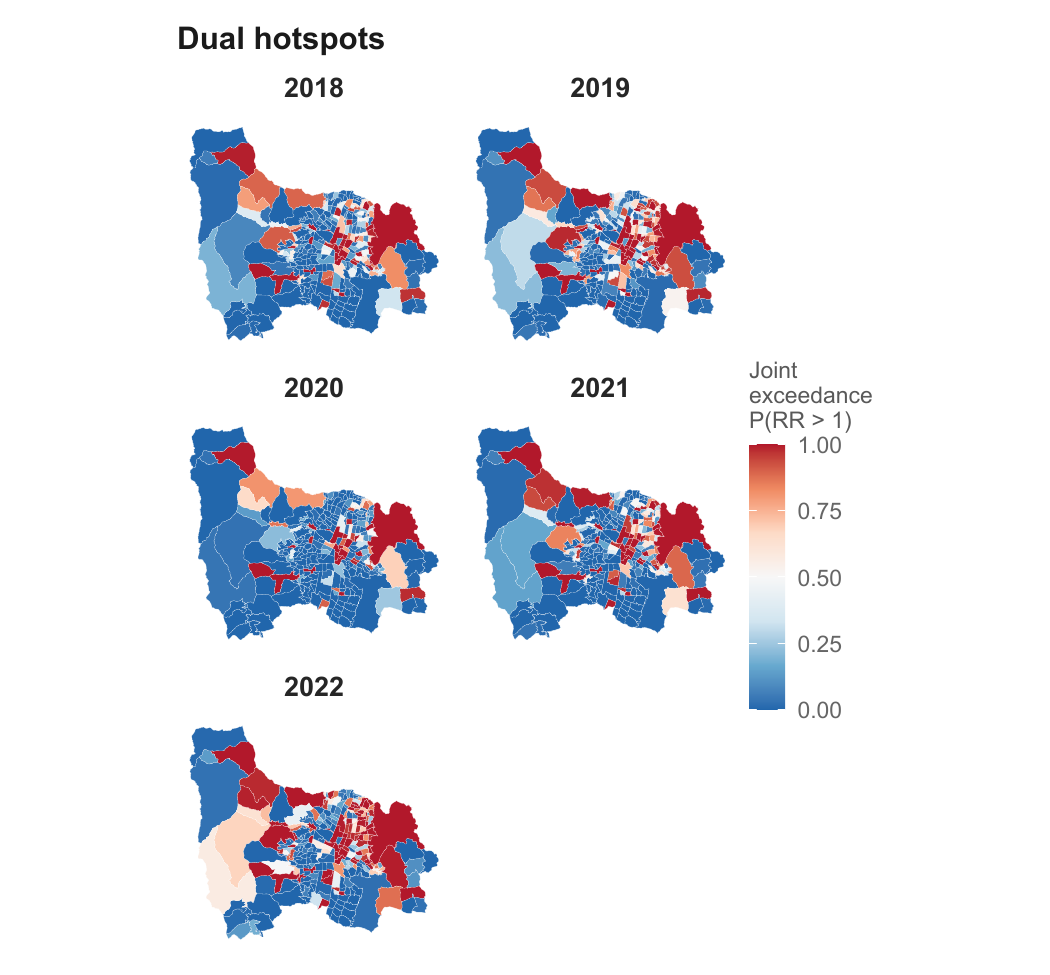}
    \caption{Joint posterior probability that both notification relative risks exceed one, by neighbourhood and year. Reference rates are outcome-specific and age-specific, pooled over 2018--2022.}
    \label{fig:hotspots}
\end{figure*}

The numbers of neighbourhoods with $p_{it}>0.8$ were 50, 72, 36, 60 and 94 in 2018--2022, respectively (Fig~\ref{fig:hotspots}). Because the reference rates were pooled over the study period, these annual counts reflect changes in city-wide notification levels as well as the spatial distribution and uncertainty of relative risks.

Twenty-five neighbourhoods met the persistence criterion of $p_{it}>0.95$ in every year. Six were in La Candelaria, seven across Manrique, Aranjuez, Villa Hermosa and Buenos Aires, five in rural districts and seven elsewhere. The threshold applies separately to each year; it does not imply a posterior probability above $0.95$ of simultaneous exceedance across all five years. This classification remains conditional on the fitted model, whose time-constant spatial fields favour recurring spatial patterns.

Small resident populations affected interpretation. Nine of the 25 persistent hotspots had fewer than 500 resident women, compared with a median of $3{,}790$ across all 285 analysis units. Several were in the city centre or largely non-residential areas, where notifications may involve people not represented by resident-population denominators. For example, the El Poblado neighbourhood had 224 resident women and mean estimated relative risks of $11.5$ for non-sexual and $23.5$ for sexual violence.

These results identify areas for closer review of notification patterns and local service context. They should not be interpreted as a ranking of residents' individual risk or as evidence of where violence occurred, particularly in areas serving substantial non-resident populations.

\FloatBarrier
\section{Discussion}\label{sec:discussion}

Across the 285 neighbourhoods of Medellín, the shared spatial component contributed more strongly to the gradients of both sexual and non-sexual violence notifications during the 2020--2021 restriction period, without clear evidence of a change in their relative loading. In 2022, the non-sexual loading remained above its initial value, whereas the sexual-violence loading was estimated close to its pre-pandemic level.

These changes coincided with a period in which reviews documented disruptions in violence and reporting, including increases in reported domestic and intimate-partner violence during stay-at-home orders \citep{Evans2020,Piquero2021}. Our findings concern the amplitude of a shared spatial pattern in notifications. Its stronger contribution to spatial contrasts coexisted with a city-wide decline in notifications in 2020 (Section~\ref{sec:posterior}): lower overall notification levels can occur alongside a steeper gradient contributed by the shared component. Changes in violence, access to services and notification practices could all contribute to this pattern, and the data do not distinguish their respective roles.

The shared field should therefore be interpreted as reflecting both the contexts in which violence is experienced and the processes through which it is notified. At commune level, it was generally higher where living conditions were poorer, as in the north-eastern communes (Section~\ref{sec:decomposition}). La Candelaria was an exception, combining relatively favourable living conditions with the highest shared-field values. Its commercial activity and large non-resident daytime population \citep{MedellinCandelariaPlan} raise questions about the correspondence between notifications and resident-population denominators. Beyond the shared pattern, non-sexual violence retained a smaller additional component concentrated in peripheral and largely rural areas.

The aggregate findings persisted when the analysis was restricted to notifications from health institutions: both loadings increased during restrictions, and the non-sexual loading remained elevated in 2022. The main pattern was thus retained after excluding family commissary notifications, which showed the principal documented change in age composition. The exploratory analyses by perpetrator relationship were more sensitive to the reporting source. Opposing patterns across relationship strata in the full dataset were not retained when each stratum was restricted to health-institution notifications. These results support distinguishing the aggregate findings from the less stable stratified patterns; they do not rule out changes in access or detection within health institutions, nor establish that events moved into homes.

The modelling approach builds on established shared-component disease-mapping models \citep{KnorrHeldBest2001,Held2005} and formulations with outcome- and time-specific spatial loadings \citep{Baker2017}. Allowing both loadings to vary after a reference period made it possible to distinguish a common steepening from a change in their relative contribution. This distinction changed the interpretation obtained under M3, where the non-sexual loading was fixed throughout. Because M4 and M3 differed by less than one standard error of WAIC, the choice of M4 rested primarily on its ability to represent the quantities of interest, rather than a clear predictive advantage. Combining these contrasts with age-standardised expected counts and joint exceedance mapping extends the application of shared-component models to an intraurban setting \citep{VicenteSER2020}.

This interpretation remains conditional on the model structure. The shared field is assumed constant over time, so changing loadings describe changes in its contribution rather than unrestricted changes in the underlying geography. SCM provides an alternative representation through a time-varying loading on a shared space-time interaction \citep{Retegui2024}. Its comparison with M3 and M4 does not isolate the effect of a single component because their spatial and temporal specifications also differ. Moreover, the levels of the sexual loadings trade off against the weakly identified sexual-specific field. The period contrasts were stable across the starting values and field scales examined, supporting their use as the main inferential summaries without resolving the identification of individual components. Changes in these loadings do not, by themselves, imply changes in the correlation between the full risk surfaces.

Several limitations affect interpretation. The data comprise suspected cases that came into contact with health or protection services, with potentially substantial and differential underreporting, particularly for sexual violence \citep{WHOSexualViolenceSurveillance}. Restricting the analysis to health-institution notifications does not eliminate changes in access or ascertainment during restrictions, and age standardisation cannot correct such changes. The five-year series also provides limited information on the pre-pandemic trend. The period contrasts therefore describe temporal associations rather than causal effects of restrictions. In addition, predictive checks identified a departure for non-sexual notifications whose influence on the loading contrasts was not directly quantified.

Finally, the model includes no neighbourhood-level covariates, so its latent fields describe spatial variation without identifying the characteristics that generate it. Adding contextual covariates would require attention to spatial confounding, because covariates with smooth spatial patterns may overlap with structured fields and affect estimated effects and their uncertainty \citep{Adin2023}. The estimated loading changes do not establish that population composition or unmeasured contextual factors are irrelevant.

For local public-health teams, joint exceedance maps could help identify neighbourhoods where elevated notifications of both outcomes warrant closer review. These maps should be interpreted alongside population denominators, the full relative-risk estimates and their uncertainty. Teams could compare the patterns with information on age-appropriate prevention, referral and care for girls, adolescents and adults. Because service coverage and barriers to access were not measured, elevated notification risks do not establish a service gap. Areas with few notifications also merit attention, as limited reporting access may obscure need.

Comparison with forensic records from the National Institute of Legal Medicine could help assess whether the patterns extend across surveillance sources, accounting for differences in case definitions, coverage and access. Such comparisons would not, by themselves, quantify all unreported violence. Together with service-coverage information, they could help evaluate the usefulness of these maps for local planning. Applications in other cities should likewise consider their reporting systems, geographic units and population references.

\section{Conclusion}\label{sec:conclusion}

The joint analysis of sexual and non-sexual violence notifications in Medellín during 2018--2022 identified a shared neighbourhood pattern whose contribution to both outcomes increased during the restriction period, without clear evidence of a change in their relative loading. In 2022, the non-sexual loading remained elevated, while the sexual-violence loading was estimated close to its initial value. These patterns were retained among health-institution notifications, although the period contrasts do not identify a causal effect of restrictions. Joint exceedance maps could support local review of surveillance and access to prevention and care, with attention to population denominators and uncertainty in the estimated risks.

\section{Source Code}

The R code for data preparation, model fitting, sensitivity analyses, and the tables and figures is available at \url{https://github.com/torodriguezt/medellin-violence-mapping}. The repository contains one script per model, together with the population projections and neighbourhood cartography used in the analysis. Case-level notification records are excluded because they contain sensitive health information; the required input file and variables are documented in the repository. The analysis used R 4.4.2 and R-INLA 24.12.11.

\clearpage
\appendix

\section{Model specifications and additional results}
\label{app:models}

\subsection{Model formulations}

All models use the Poisson observation model in Eq~\eqref{eq:pois_st}, where $\eta_{ikt}$ is the log-relative risk for outcome $k$ in neighbourhood $i$ and year $t$, with non-sexual violence as the reference outcome ($k=1$). The BYM2 fields and temporal random walks follow the specifications in the main text.

\paragraph{M0: separable baseline.}
\begin{equation*}
\eta_{ikt} = \alpha_k + \theta_{ik} + \beta_{kt},
\end{equation*}
where $(\theta_{i1},\theta_{i2})$ is a BYM2 spatial field replicated over the two outcomes with an exchangeable between-outcome correlation, and $(\beta_{1t},\beta_{2t})$ an RW2 temporal trend replicated in the same way. The model has no explicit common component or space--time interaction.

\paragraph{M1: static shared components.}
\begin{align*}
\eta_{i1t} &= \alpha_1 + \psi_i + \phi_{i1} + \gamma_t,\\
\eta_{i2t} &= \alpha_2 + \delta\,\psi_i + \phi_{i2}
             + \lambda\,\gamma_t + \nu_t,
\end{align*}
with $\psi_i$, $\phi_{i1}$ and $\phi_{i2}$ BYM2 fields, $\gamma_t$ an RW2, $\nu_t$ an RW1, and $\delta$ and $\lambda$ scalar loadings.

\paragraph{M2-I--M2-IV: outcome-specific interactions.}
\begin{align*}
\eta_{i1t} &= \alpha_1 + \psi_i + \phi_{i1}
             + \gamma_t + \Delta_{i1t},\\
\eta_{i2t} &= \alpha_2 + \delta\,\psi_i + \phi_{i2}
             + \lambda\,\gamma_t + \nu_t + \Delta_{i2t},
\end{align*}
where the interaction of outcome $k$ has precision matrix $\tau_k\mathbf{R}$, with
\begin{equation*}
\mathbf{R} =
\begin{cases}
\mathbf{I}_T\otimes\mathbf{I}_A, & \text{Type I},\\
\mathbf{R}_T\otimes\mathbf{I}_A, & \text{Type II},\\
\mathbf{I}_T\otimes\mathbf{R}_S, & \text{Type III},\\
\mathbf{R}_T\otimes\mathbf{R}_S, & \text{Type IV}.
\end{cases}
\end{equation*}
Here $\mathbf{R}_T$ is the RW2 structure matrix on the years and $\mathbf{R}_S$ the intrinsic CAR structure on the neighbourhood graph.

\paragraph{M3: time-varying relative spatial loading.}
M3 retains the Type~I interaction and replaces the constant spatial loading of M2-I with a time-varying loading:
\begin{align*}
\eta_{i1t} &= \alpha_1 + \psi_i + \phi_{i1}
             + \gamma_t + \Delta_{i1t},\\
\eta_{i2t} &= \alpha_2 + \delta_t\,\psi_i + \phi_{i2}
             + \lambda\,\gamma_t + \nu_t + \Delta_{i2t},\\
\delta_t &= \sum_{l=1}^{L}\delta_l\,\mathbf{1}\{t\in P_l\},
\end{align*}
where the $L$ periods $P_l$ are either the five individual years ($L=T$) or the three periods in Eq~\eqref{eq:l3} ($L=3$).

\paragraph{M4: period-specific loadings in both outcomes.}
M4 extends three-period M3 by allowing the non-sexual loading to vary between periods:
\begin{align*}
\eta_{i1t} &= \alpha_1 + \delta_{1t}\,\psi_i + \phi_{i1}
             + \gamma_t + \Delta_{i1t},\\
\eta_{i2t} &= \alpha_2 + \delta_{2t}\,\psi_i + \phi_{i2}
             + \lambda\,\gamma_t + \nu_t + \Delta_{i2t},\\
\delta_{kt} &= \sum_{l=1}^{3}\delta_{kl}\,\mathbf{1}\{t\in P_l\},
\qquad \delta_{1,\text{pre}}=1,
\end{align*}
with the periods $P_l$ of Eq~\eqref{eq:l3}. The five remaining spatial loadings are free. Three-period M3 is the special case $\delta_{1t}=1$ for all $t$.

\paragraph{SCM: time-varying shared-interaction loadings.}
The alternative model motivated by \citet{Retegui2024} is
\begin{align*}
\eta_{i1t} &= \alpha_1 + \kappa u_i + \beta_{1t}
             + \varrho_t\,\Delta^{\text{sh}}_{it},\\
\eta_{i2t} &= \alpha_2 + \kappa^{-1}u_i + v_i + \beta_{2t}
             + \varrho_t^{-1}\,\Delta^{\text{sh}}_{it},
\end{align*}
with $u_i$ an intrinsic CAR field shared by the two outcomes through reciprocal static scalings $\kappa$ and $\kappa^{-1}$, $v_i$ unstructured heterogeneity for sexual violence, $\beta_{kt}$ an outcome-specific RW1 trend, and $\Delta^{\text{sh}}_{it}$ a single Type~I space--time interaction shared by both outcomes and loaded through the reciprocal year-specific scalings $\varrho_t$ and $\varrho_t^{-1}$. Under this parameterisation, $1/\varrho_t^{2}$ is the relative interaction loading for sexual violence compared with non-sexual violence.

\subsection{Prior distributions}

Table~\ref{tab:priors} summarises the priors for the BYM2 shared-component models M1--M4. The normal priors apply only to free loadings.

\begin{table}[pos=htbp]
\centering
\small
\caption{Components and priors of the BYM2 shared-component models M1--M4. Fixed reference loadings are excluded from the loading priors.}
\label{tab:priors}
\begin{tabular*}{\tblwidth}{@{}LLL@{}}
\toprule
Component & Form & Prior \\
\midrule
Shared spatial $\psi_i$
    & BYM2
    & \makecell[l]{$P(\sigma>1)=0.01$\\
                   $P(\pi<0.5)=\tfrac{2}{3}$} \\
Specific spatial $\phi_{ik}$
    & BYM2
    & as above \\
Shared trend $\gamma_t$
    & RW2
    & $P(\sigma>1)=0.01$ \\
Sexual deviation $\nu_t$
    & RW1
    & $P(\sigma>1)=0.01$ \\
Interaction $\Delta_{ikt}$
    & Knorr-Held I--IV
    & $P(\sigma>1)=0.01$ \\
\addlinespace
\makecell[l]{Free spatial loadings\\
             $\delta$, $\delta_t$, $\delta_{kt}$}
    & Scalar
    & $\mathcal{N}(0,1)$ \\
Temporal loading $\lambda$
    & Scalar
    & $\mathcal{N}(0,1)$ \\
\bottomrule
\end{tabular*}
\end{table}

For SCM, the priors are $\kappa\sim\mathrm{Gamma}(10,10)$, independent $\varrho_t\sim\mathrm{Gamma}(10,10)$, and $\tau_{\mathrm{sh}}\sim\mathrm{Gamma}(1,5\times10^{-5})$ for the shared-interaction precision, with Gamma distributions parameterised by shape and rate. The spatial field, sexual-specific heterogeneity and both RW1 trends use improper flat priors on their standard-deviation scales, corresponding to $p(\tau)\propto\tau^{-3/2}$. The prior on the relative interaction loading $1/\varrho_t^{2}$ is induced by this transformation.

\subsection{Constraints and implementation}

Intrinsic components were constrained to sum to zero. For the outcome-specific interactions, the additional centring constraints and diagonal regularisation used for each Knorr-Held type are detailed in Section~\ref{sec:interactions}, with constraints informed by \citet{Goicoa2018}.

In M3 and M4, the effective precision of the shared temporal component was
\begin{equation*}
\mathbf{Q}_\gamma
=
\tau_\gamma c\mathbf{R}_2
+
10^{-4}\mathbf{I}_T,
\end{equation*}
subject to $\sum_t\gamma_t=0$, where $\mathbf{R}_2$ is the second-order difference structure matrix and $c\approx0.2775821$ is its scaling factor for the five years. No additional constraint removed the linear component.

Spatial and temporal loadings in M1--M4 were implemented using INLA's \texttt{copy} feature. SCM used an \texttt{rgeneric} construction for its shared interaction with time-varying loadings. Its equations describe the structural representation; the implementation uses numerical copy approximations and diagonal regularisation. The shared interaction is constrained to sum to zero separately over all areas and years within each loaded outcome-specific block.

\subsection{Interpretation of loading contrasts}

For two neighbourhoods $i$ and $j$, in period $t$, the shared component contributes $\delta_{kt}(\psi_i-\psi_j)$ to their log-risk difference for outcome $k$, or $\exp[\delta_{kt}(\psi_i-\psi_j)]$ to their risk ratio. The change in this contribution between the pre-pandemic and restriction periods is
\begin{equation}
\frac{\exp[\delta_{k,\text{res}}(\psi_i-\psi_j)]}
     {\exp[\delta_{k,\text{pre}}(\psi_i-\psi_j)]}
= \exp\!\left[D_k(\psi_i-\psi_j)\right].
\label{eq:delta_pair}
\end{equation}

For $D_k>0$, this factor exceeds one when $\psi_i>\psi_j$, equals one when the shared-field values are equal, and is below one when $\psi_i<\psi_j$. Its magnitude therefore depends on the neighbourhood pair; $D_k$ alone is not a percentage change in risk. Uncertainty must be propagated by evaluating Eq~\eqref{eq:delta_pair} within each joint posterior draw of the loadings and shared field.

\subsection{Model comparison and computing times}

Let $\boldsymbol{\theta}$ denote the latent field and hyperparameters, and index the $N=2AT$ observations by $j$. The deviance information criterion \citep{spiegelhalter2002} is
\begin{equation}
\begin{gathered}
D(\boldsymbol{\theta})
= -2\log p(\mathbf{y}\mid\boldsymbol{\theta}),\\[4pt]
\overline{D}
= \mathbb{E}_{\boldsymbol{\theta}\mid\mathbf{y}}
  [D(\boldsymbol{\theta})],\\[4pt]
p_D = \overline{D}-D(\bar{\boldsymbol{\theta}}),
\qquad
\text{DIC} = \overline{D}+p_D.
\end{gathered}
\label{eq:dic}
\end{equation}
Here $\overline{D}$ is the posterior mean deviance, $\bar{\boldsymbol{\theta}}$ the posterior mean of $\boldsymbol{\theta}$, and $p_D$ the effective number of parameters.

The Watanabe--Akaike information criterion \citep{watanabe2010} is computed pointwise:
\begin{equation}
\begin{split}
\text{WAIC} ={}&
-2\sum_j
\log \mathbb{E}_{\boldsymbol{\theta}\mid\mathbf{y}}
\!\left[p(y_j\mid\boldsymbol{\theta})\right]\\
&+
2\sum_j
\operatorname{Var}_{\boldsymbol{\theta}\mid\mathbf{y}}
\!\left[\log p(y_j\mid\boldsymbol{\theta})\right].
\end{split}
\label{eq:waic}
\end{equation}
The second term equals twice the WAIC effective number of parameters.

Table~\ref{tab:cpu} reports the total elapsed times recorded by INLA, including preprocessing, fitting and postprocessing. All models used \texttt{2:1} threads, sometimes alongside other fits, so these times describe the saved runs rather than a controlled comparison of computational efficiency.

\begin{table}[pos=htbp]
\centering
\small
\caption{Total elapsed time recorded by INLA for the fits.}
\label{tab:cpu}
\begin{tabular*}{\tblwidth}{@{}LL@{}}
\toprule
Model & Elapsed time (s) \\
\midrule
M0: separable baseline & 10.2 \\
M1: static shared components & 41.6 \\
M2-I: Type I interaction & 86.1 \\
M2-III: Type III interaction & 251.9 \\
M3: annual $\delta_t$ ($L=T$) & 143.5 \\
M3: three-period $\delta_t$ ($L=3$) & 93.2 \\
M4: three-period loadings in both outcomes & 133.4 \\
SCM: shared interaction $\varrho_t$ & 47.6 \\
\bottomrule
\end{tabular*}
\end{table}

\subsection{Identifiability and sensitivity analyses}

The levels of the sexual loadings depend on the allocation of spatial variation between the shared and sexual-specific fields. For any common constant $c$, replacing $\delta_{2t}$ with $\delta_{2t}+c$ and $\phi_{i2}$ with $\phi_{i2}-c\,\psi_i$ leaves the predictor contribution $\delta_{2t}\psi_i+\phi_{i2}$ unchanged. The likelihood alone therefore does not identify the level of the sexual loadings; their levels depend on the priors for the loadings and $\phi_2$.

Under this transformation, $D_2$ is unchanged, whereas the relative-loading contrast becomes
\begin{equation*}
D_\rho+c\left(\frac{1}{\delta_{1,\text{res}}}-1\right).
\end{equation*}
The stability of $D_\rho$ across fits is thus an empirical result, not a consequence of expressing it as a difference.

\paragraph{Starting values and sensitivity to the sexual-specific scale.}

We examined starting values of $0$, $0.5$, $1$, $1.5$ and $-1$ for the free loadings. A solution was considered stable when restarting the mode search reproduced it. The reported fit, obtained from a starting value of $0$, met this criterion: restarting from that fit, or starting from $0.5$ with a restart, reproduced the solution (Table~\ref{tab:appA_starts}).

Without a restart, starting values of $0.5$ and $1$ produced solutions that subsequently changed when the search was restarted. We therefore classified these as incomplete mode searches. Their WAIC values were higher than those of the reported fit. The slightly higher log marginal likelihood obtained from $1$ (about $0.5$ units) was close to the variation between repeated fits reaching the same solution (about $0.3$ units).

Starting from $1$ or $1.5$ with a restart reached a second stable solution, with a sexual-specific scale of about $0.07$ rather than $0.21$, a log marginal likelihood 1--2 units lower, and slightly smaller non-sexual loading changes. These solutions are consistent with weak identification of the sexual-specific spatial decomposition. A starting value of $-1$ reached a mode with negative loadings after 2018--2019 and a log marginal likelihood about $120$ units lower.

Fixing the scale of $\phi_2$ at values between $0.001$ and $0.3$ provided a further assessment of this dependence. As the fixed scale increased, the sexual loadings decreased but the period contrasts changed little. Within this comparison, log marginal likelihoods varied by about 2 units and were highest at a scale of $0.2$, close to the estimated scale in the reported fit. Removing $\phi_2$ also produced similar period contrasts. We retained this field to allow a separate time-constant spatial main effect for sexual violence.

\begin{table*}[pos=tp]
\centering
\footnotesize
\caption{Sensitivity of M4 to loading starting values, mode-search restarts, and the scale or inclusion of the sexual-specific field $\phi_2$. Contrasts compare the indicated period with 2018--2019 and are posterior means (95\% credible intervals) from $5{,}000$ joint hyperparameter draws per fit.}
\label{tab:appA_starts}
\begin{tabular*}{\tblwidth}{@{}LLLLLL@{}}
\toprule
Fit & WAIC
& \makecell[l]{Non-sexual\\2020--2021}
& \makecell[l]{Sexual\\2020--2021}
& \makecell[l]{Relative loading\\2020--2021}
& \makecell[l]{Non-sexual\\2022} \\
\midrule
\multicolumn{6}{@{}l}{\textit{Starting value, without restart}} \\
0 (reported fit) & 15{,}253 & $+0.26$ (0.16, 0.37) & $+0.23$ (0.09, 0.37) & $-0.06$ ($-0.21$, 0.10) & $+0.19$ (0.07, 0.31) \\
0, refitted & 15{,}252 & $+0.26$ (0.15, 0.37) & $+0.23$ (0.09, 0.37) & $-0.06$ ($-0.21$, 0.09) & $+0.19$ (0.07, 0.31) \\
0.5 & 15{,}263 & $+0.23$ (0.13, 0.34) & $+0.24$ (0.11, 0.38) & $-0.05$ ($-0.20$, 0.11) & $+0.16$ (0.04, 0.28) \\
1 & 15{,}260 & $+0.19$ (0.03, 0.33) & $+0.23$ (0.10, 0.37) & $-0.02$ ($-0.19$, 0.19) & $+0.12$ ($-0.03$, 0.26) \\
$-1$ & 15{,}264 & \multicolumn{4}{l}{negative loadings after 2018--2019} \\
\addlinespace
\multicolumn{6}{@{}l}{\textit{Starting value, with restart}} \\
0 & 15{,}255 & $+0.26$ (0.16, 0.37) & $+0.23$ (0.09, 0.37) & $-0.06$ ($-0.22$, 0.09) & $+0.19$ (0.07, 0.31) \\
0.5 & 15{,}252 & $+0.26$ (0.15, 0.37) & $+0.23$ (0.09, 0.38) & $-0.06$ ($-0.21$, 0.09) & $+0.19$ (0.07, 0.31) \\
1 & 15{,}253 & $+0.24$ (0.14, 0.35) & $+0.24$ (0.11, 0.38) & $-0.05$ ($-0.20$, 0.10) & $+0.17$ (0.06, 0.29) \\
1, refitted & 15{,}256 & $+0.24$ (0.14, 0.35) & $+0.24$ (0.11, 0.38) & $-0.05$ ($-0.21$, 0.10) & $+0.17$ (0.05, 0.29) \\
1.5 & 15{,}254 & $+0.24$ (0.14, 0.35) & $+0.24$ (0.11, 0.38) & $-0.05$ ($-0.20$, 0.10) & $+0.17$ (0.06, 0.29) \\
\addlinespace
\multicolumn{6}{@{}l}{\textit{Scale of $\phi_2$ fixed (starting value 0)}} \\
0.001 & 15{,}252 & $+0.24$ (0.14, 0.34) & $+0.24$ (0.11, 0.37) & $-0.05$ ($-0.20$, 0.10) & $+0.17$ (0.05, 0.28) \\
0.05 & 15{,}254 & $+0.24$ (0.14, 0.34) & $+0.24$ (0.11, 0.37) & $-0.05$ ($-0.20$, 0.11) & $+0.17$ (0.06, 0.29) \\
0.1 & 15{,}253 & $+0.25$ (0.15, 0.36) & $+0.24$ (0.11, 0.38) & $-0.06$ ($-0.21$, 0.10) & $+0.18$ (0.06, 0.30) \\
0.2 & 15{,}252 & $+0.26$ (0.16, 0.37) & $+0.24$ (0.09, 0.37) & $-0.06$ ($-0.21$, 0.10) & $+0.19$ (0.07, 0.31) \\
0.3 & 15{,}251 & $+0.23$ (0.12, 0.34) & $+0.19$ (0.05, 0.33) & $-0.06$ ($-0.20$, 0.09) & $+0.17$ (0.05, 0.28) \\
\addlinespace
Without $\phi_2$ (all starting values) & 15{,}257 & $+0.24$ (0.14, 0.35) & $+0.24$ (0.11, 0.37) & $-0.05$ ($-0.20$, 0.10) & $+0.17$ (0.05, 0.28) \\
\bottomrule
\end{tabular*}
\end{table*}
\paragraph{Alternative specifications and data subsets.}

Table~\ref{tab:appA_m4_sens} summarises the period contrasts under alternative loading priors, spatial structures and interaction types, followed by changes in notification selection and exploratory analyses by perpetrator relationship.

Among the sensitivity analyses reported in Table~\ref{tab:appA_m4_sens}, contrasts from non-negative starting values agreed within $0.04$, except for the Type~III interaction specification, which reached a negative-loading mode when started at $0$. We report its fit started at $1$, whose log marginal likelihood was $106$ units higher.

\begin{table*}[pos=!tp]
\centering
\footnotesize
\caption{M4 sensitivity analyses: changes in spatial loadings and relative loading ($\rho=\delta_2/\delta_1$) from 2018--2019 to the indicated period. Entries are posterior means (95\% credible intervals) from $5{,}000$ joint hyperparameter draws per fit.}
\label{tab:appA_m4_sens}
\begin{tabular*}{\tblwidth}{@{}LLLLL@{}}
\toprule
Analysis
& \makecell[l]{Non-sexual\\2020--2021}
& \makecell[l]{Sexual\\2020--2021}
& \makecell[l]{Relative loading\\2020--2021}
& \makecell[l]{Non-sexual\\2022} \\
\midrule
All notifications (main) & $+0.26$ (0.16, 0.37) & $+0.23$ (0.09, 0.37) & $-0.06$ ($-0.21$, 0.10) & $+0.19$ (0.07, 0.31) \\
\addlinespace
\multicolumn{5}{@{}l}{\textit{Model specification}} \\
Loading prior $\mathcal{N}(0,4)$ & $+0.27$ (0.16, 0.38) & $+0.24$ (0.09, 0.38) & $-0.07$ ($-0.22$, 0.09) & $+0.20$ (0.08, 0.32) \\
Loading prior $\mathcal{N}(1,1)$ & $+0.27$ (0.17, 0.38) & $+0.24$ (0.10, 0.38) & $-0.07$ ($-0.23$, 0.09) & $+0.20$ (0.08, 0.32) \\
Intrinsic CAR spatial fields & $+0.25$ (0.15, 0.36) & $+0.23$ (0.09, 0.37) & $-0.06$ ($-0.21$, 0.10) & $+0.19$ (0.08, 0.30) \\
Type~III interaction & $+0.28$ (0.17, 0.38) & $+0.26$ (0.12, 0.40) & $-0.07$ ($-0.23$, 0.09) & $+0.15$ (0.04, 0.27) \\
\addlinespace
\multicolumn{5}{@{}l}{\textit{Notification selection}} \\
Health institutions only & $+0.30$ (0.18, 0.41) & $+0.22$ (0.09, 0.35) & $-0.08$ ($-0.21$, 0.06) & $+0.18$ (0.05, 0.30) \\
Women aged 20 and over & $+0.24$ (0.12, 0.36) & $+0.30$ (0.06, 0.54) & $+0.05$ ($-0.18$, 0.29) & $+0.20$ (0.07, 0.33) \\
Without neglect & $+0.29$ (0.17, 0.41) & $+0.24$ (0.10, 0.39) & $-0.08$ ($-0.24$, 0.08) & $+0.21$ (0.08, 0.34) \\
\addlinespace
\multicolumn{5}{@{}l}{\textit{Exploratory analyses by perpetrator relationship}} \\
Intrafamilial & $+0.32$ (0.20, 0.43) & $+0.06$ ($-0.13$, 0.25) & $-0.29$ ($-0.48$, $-0.09$) & $+0.08$ ($-0.04$, 0.20) \\
Other relationships & $-0.14$ ($-0.29$, 0.02) & $+0.15$ (0.02, 0.27) & $+0.35$ (0.10, 0.64) & $+0.43$ (0.15, 0.71) \\
Intrafamilial, health only & $+0.21$ (0.11, 0.32) & $+0.06$ ($-0.09$, 0.20) & $-0.13$ ($-0.28$, 0.02) & $+0.14$ (0.02, 0.26) \\
Other relationships, health only & $+0.41$ (0.14, 0.68) & $+0.22$ (0.07, 0.38) & $-0.15$ ($-0.39$, 0.10) & $+0.25$ ($-0.05$, 0.56) \\
\bottomrule
\end{tabular*}
\par\smallskip
\begin{minipage}{\tblwidth}
\footnotesize
\textit{Note.} Expected counts were recomputed for each data subset; relationship strata use crude outcome-specific reference rates. Each analysis estimates its own shared field, so loading changes are expressed in that field's units; $\rho$ is unit-free. Health-only analyses exclude family commissary notifications. The Type~III interaction fit used loading starting values of 1.
\end{minipage}
\end{table*}

\subsection{Additional posterior and spatial summaries}

Table~\ref{tab:appA_hyper_full} reports posterior summaries for M4 from INLA's integrated marginal approximations. The ratios and period contrasts in Table~\ref{tab:period_contrasts} instead use joint draws to retain dependence between loadings. For random walks, the scales govern increment penalties; for constrained interactions, they describe the underlying independent prior before conditioning. The scale of $\phi_2$ and the levels of the sexual loadings are weakly identified (Table~\ref{tab:appA_starts}).

\begin{table*}[pos=!tp]
\centering
\small
\caption{M4 posterior means and 95\% credible intervals from INLA's integrated marginals. Scales are summarised as $\sigma=\tau^{-1/2}$; $\pi$ is the nominal structured share of spatial variance.}
\label{tab:appA_hyper_full}
\begin{tabular*}{\tblwidth}{@{}LLL@{}}
\toprule
Parameter & Description & Posterior mean (95\% CrI) \\
\midrule
$\alpha_1$ & Intercept, non-sexual violence & $-0.12$ ($-0.18$, $-0.06$) \\
$\alpha_2$ & Intercept, sexual violence & $-0.15$ ($-0.22$, $-0.09$) \\
\addlinespace
$\sigma_{\psi}$ & Shared spatial field, scale & 0.50 (0.44, 0.56) \\
$\pi_{\psi}$ & Shared spatial field, structured proportion & 0.36 (0.19, 0.55) \\
$\sigma_{\phi_1}$ & Non-sexual-specific spatial field, scale & 0.15 (0.08, 0.24) \\
$\pi_{\phi_1}$ & Non-sexual-specific spatial field, structured proportion & 0.42 (0.02, 0.89) \\
$\sigma_{\phi_2}$ & Sexual-specific spatial field, scale & 0.21 (0.14, 0.32) \\
$\pi_{\phi_2}$ & Sexual-specific spatial field, structured proportion & 0.89 (0.63, 0.99) \\
\addlinespace
$\delta_{1,\text{res}}$ & Non-sexual spatial loading, 2020--2021 & 1.26 (1.16, 1.37) \\
$\delta_{1,\text{post}}$ & Non-sexual spatial loading, 2022 & 1.19 (1.07, 1.31) \\
$\delta_{2,\text{pre}}$ & Sexual spatial loading, 2018--2019 & 1.18 (1.03, 1.34) \\
$\delta_{2,\text{res}}$ & Sexual spatial loading, 2020--2021 & 1.41 (1.23, 1.59) \\
$\delta_{2,\text{post}}$ & Sexual spatial loading, 2022 & 1.20 (1.03, 1.37) \\
\addlinespace
$\lambda$ & Temporal loading, sexual violence & 0.83 (0.15, 1.50) \\
$\sigma_\gamma$ & Shared RW2 scale & 0.28 (0.13, 0.54) \\
$\sigma_\nu$ & Sexual-specific RW1 scale & 0.17 (0.08, 0.32) \\
\addlinespace
$\sigma_{\Delta_1}$ & Non-sexual interaction, prior scale & 0.19 (0.17, 0.21) \\
$\sigma_{\Delta_2}$ & Sexual interaction, prior scale & 0.18 (0.15, 0.21) \\
\bottomrule
\end{tabular*}
\end{table*}

Table~\ref{tab:appA_m3_periods} presents the spatial loadings and contrasts for the three-period (panel A) and annual (panel B) versions of M3.

\begin{table*}[pos=!tp]
\centering
\small
\caption{M3 spatial loadings and contrasts for (A) the three-period and (B) the annual specification, with the non-sexual loading fixed at one. Entries are posterior means (95\% credible intervals). Loading summaries use INLA's integrated marginals; contrasts and probabilities use $5{,}000$ joint hyperparameter draws. In panel B, $\bar{\delta}_{\mathrm{pre}}=(\delta_{2018}+\delta_{2019})/2$; dashes mark the reference years.}
\label{tab:appA_m3_periods}
\label{tab:annual_loadings}
\begin{tabular*}{\tblwidth}{@{}LLL@{}}
\toprule
\multicolumn{3}{@{}l}{\textit{A. Three-period specification}} \\
\addlinespace
Quantity & Mean (95\% CrI) & Probability \\
\midrule
$\delta_{\text{pre}}$ (2018--2019) & 1.16 (1.03, 1.33) & --- \\
$\delta_{\text{res}}$ (2020--2021) & 1.36 (1.22, 1.53) & --- \\
$\delta_{\text{post}}$ (2022) & 1.16 (1.03, 1.33) & --- \\
\addlinespace
$\delta_{\text{res}}-\delta_{\text{pre}}$ & $+0.20$ (0.08, 0.32) & $\Pr(>0)=0.999$ \\
$\delta_{\text{post}}-\delta_{\text{res}}$ & $-0.20$ ($-0.33$, $-0.07$) & $\Pr(<0)=0.998$ \\
$\delta_{\text{post}}-\delta_{\text{pre}}$ & $+0.00$ ($-0.13$, 0.13) & $\Pr(>0)=0.526$ \\
\bottomrule
\end{tabular*}

\par\medskip

\begin{tabular*}{\tblwidth}{@{}LLLL@{}}
\toprule
\multicolumn{4}{@{}l}{\textit{B. Annual specification}} \\
\addlinespace
Year & $\delta_t$ (95\% CrI) & $\delta_t-\bar{\delta}_{\mathrm{pre}}$ (95\% CrI) & $\Pr(\delta_t>\bar{\delta}_{\mathrm{pre}})$ \\
\midrule
2018 & 0.964 (0.823, 1.104) & --- & --- \\
2019 & 1.052 (0.917, 1.187) & --- & --- \\
2020 & 1.233 (1.075, 1.389) & 0.223 (0.075, 0.375) & 0.998 \\
2021 & 1.138 (0.992, 1.282) & 0.127 ($-0.018$, 0.272) & 0.959 \\
2022 & 1.021 (0.893, 1.147) & 0.012 ($-0.119$, 0.142) & 0.578 \\
\bottomrule
\end{tabular*}
\end{table*}

\begin{figure*}[pos=!tp]
    \centering
    \includegraphics[width=\textwidth]{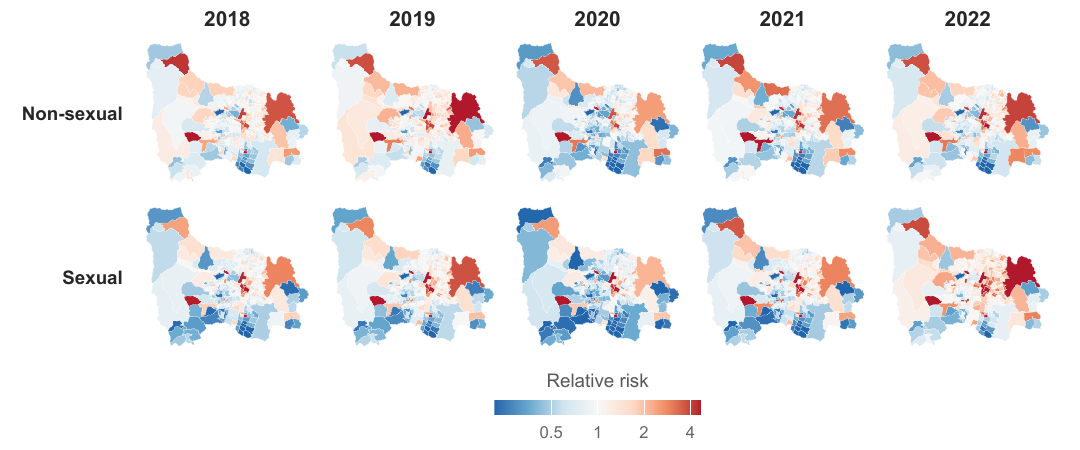}
    \caption{M4 relative risks by neighbourhood and outcome, 2018--2022. Estimates exponentiate posterior mean log-relative risks. Red indicates values above one and blue below one, relative to outcome-specific, age-specific reference rates pooled over 2018--2022. Colours use a log scale.}
    \label{fig:appA_rr_full}
\end{figure*}

Fig~\ref{fig:appA_rr_full} presents the complete five-year relative-risk maps for M4.

\FloatBarrier

\section{Joint posterior evaluation of shared-component risk contrasts}
\label{app:psi_joint}

\subsection{Paired draws of the shared field and loadings}

We used the $5{,}000$ joint hyperparameter draws underlying Table~\ref{tab:period_contrasts} as integration points. At each point, we generated one conditional Gaussian draw of the shared field using INLA's improved conditional means and the fitted constraints, preserving its pairing with all loadings.

Within draw $s$, we calculated the difference between the 90th and 10th percentiles across the 285 neighbourhoods:
\[
g^{(s)}
=
Q_{0.90}(\psi^{(s)})-Q_{0.10}(\psi^{(s)}).
\]
The shared component's contribution to the risk ratio between these positions was evaluated as $\exp(\delta_{kp}^{(s)}g^{(s)})$ for each outcome $k$ and period $p$, with $\delta_{1,\text{pre}}=1$. Following Eq~\eqref{eq:delta_pair}, its change from the pre-pandemic to the restriction period was evaluated as
\[
\exp\!\left[
\left(
\delta_{k,\text{res}}^{(s)}
-
\delta_{k,\text{pre}}^{(s)}
\right)g^{(s)}
\right].
\]

Posterior means and 95\% credible intervals summarise these derived draws. Because the percentiles were recalculated within each draw, the contrasts refer to positions in the shared field rather than a fixed pair of neighbourhoods. They describe the shared component's contribution rather than total risk.

\subsection{Sensitivity to hyperparameter integration}

Table~\ref{tab:psi_integration_sensitivity} compares these summaries with those obtained from $2{,}000$ joint posterior draws at INLA's discrete hyperparameter configurations. The discrete-configuration intervals for the non-sexual and sexual restriction-period contrasts were 15\% and 25\% narrower, respectively.

Both approaches use conditional Gaussian field draws with improved means and no skewness correction. They differ in their representation of the hyperparameter posterior; their agreement or differences do not establish exact posterior accuracy.

\begin{table*}[pos=!tp]
\centering
\small
\caption{M4 shared-component risk contrasts under continuous hyperparameter integration ($5{,}000$ draws) and discrete configurations ($2{,}000$ draws). Entries are approximate posterior means [95\% intervals] for within-draw percentile positions of the shared field. Width ratios compare continuous with discrete intervals.}
\label{tab:psi_integration_sensitivity}
\begin{tabular*}{\tblwidth}{@{}LLLL@{}}
\toprule
Quantity & Continuous integration & Discrete configurations & Width ratio \\
\midrule
$g=Q_{0.90}(\psi)-Q_{0.10}(\psi)$ & 1.16 [1.04, 1.29] & 1.17 [1.04, 1.28] & 1.03 \\
\addlinespace
\multicolumn{4}{@{}l}{\textit{Shared contributions to spatial risk ratios}} \\
Non-sexual, 2018--2019, $\exp(g)$ & 3.20 [2.84, 3.63] & 3.22 [2.84, 3.61] & 1.03 \\
Non-sexual, 2020--2021, $\exp(\delta_{1,\text{res}}g)$ & 4.34 [3.75, 5.03] & 4.38 [3.77, 5.03] & 1.01 \\
Non-sexual, 2022, $\exp(\delta_{1,\text{post}}g)$ & 3.99 [3.40, 4.67] & 4.03 [3.45, 4.65] & 1.06 \\
Sexual, 2018--2019, $\exp(\delta_{2,\text{pre}}g)$ & 3.93 [3.26, 4.74] & 3.99 [3.40, 4.70] & 1.13 \\
Sexual, 2020--2021, $\exp(\delta_{2,\text{res}}g)$ & 5.14 [4.18, 6.36] & 5.24 [4.31, 6.37] & 1.06 \\
Sexual, 2022, $\exp(\delta_{2,\text{post}}g)$ & 4.01 [3.30, 4.89] & 4.08 [3.43, 4.83] & 1.14 \\
\addlinespace
\multicolumn{4}{@{}l}{\textit{Changes relative to the pre-pandemic period}} \\
Non-sexual, restriction/pre, $\exp(D_1 g)$ & 1.36 [1.21, 1.51] & 1.36 [1.23, 1.49] & 1.18 \\
Sexual, restriction/pre, $\exp(D_2 g)$ & 1.31 [1.11, 1.54] & 1.31 [1.17, 1.49] & 1.33 \\
Non-sexual, 2022/pre & 1.25 [1.09, 1.42] & 1.25 [1.11, 1.41] & 1.08 \\
Sexual, 2022/pre & 1.02 [0.86, 1.21] & 1.02 [0.87, 1.21] & 1.05 \\
\bottomrule
\end{tabular*}
\end{table*}

\clearpage

\bibliographystyle{cas-model2-names}
\bibliography{references}

@article{riebler2016bym2,
  author  = {S{\o}rbye, Sigrunn H. and Riebler, Andrea and Simpson, Daniel and Rue, H{\aa}vard},
  title   = {An intuitive Bayesian spatial model for disease mapping that accounts for scaling},
  journal = {Statistical Methods in Medical Research},
  volume  = {25},
  number  = {4},
  pages   = {1145--1165},
  year    = {2016},
  doi     = {10.1177/0962280216660421}
}

@book{WHO2002,
  author    = {{WHO}},
  title     = {World report on violence and health},
  year      = {2002},
  publisher = {World Health Organization},
  address   = {Geneva}
}

@book{EllsbergHeise2005,
  author    = {Ellsberg, Mary and Heise, Lori},
  title     = {Researching violence against women: Practical guidelines for researchers and activists},
  year      = {2005},
  publisher = {World Health Organization},
  address   = {Geneva}
}

@article{RussoPirlott2006,
  author  = {Russo, Nancy Felipe and Pirlott, Angela},
  title   = {Gender-based violence},
  journal = {Annals of the New York Academy of Sciences},
  year    = {2006},
  volume  = {1087},
  number  = {1},
  pages   = {178--205},
  doi     = {10.1196/annals.1385.024}
}

@article{FoxBenson2006,
author = {Fox, G. L. and Benson, M. L.},
title = {Household and neighborhood contexts of intimate partner violence},
journal = {Public Health Reports},
volume = {121},
number = {4},
pages = {419--427},
year = {2006},
doi = {10.1177/003335490612100410}
}

@article{Caetano2010,
author = {Caetano, R. and Ramisetty-Mikler, S. and Harris, T. R.},
title = {Neighborhood characteristics as predictors of male to female and female to male partner violence},
journal = {Journal of Interpersonal Violence},
volume = {25},
number = {11},
pages = {1986--2009},
year = {2010},
doi = {10.1177/0886260509354497}
}

@article{Beyer2015,
  author  = {Beyer, Kirsten M. M. and Wallis, Amanda B. and Hamberger, L. Kevin},
  title   = {Neighborhood environment and intimate partner violence: A systematic review},
  journal = {Trauma, Violence, \& Abuse},
  year    = {2015},
  volume  = {16},
  number  = {1},
  pages   = {16--47},
  doi     = {10.1177/1524838013515758}
}

@article{Gracia2018,
  author  = {Gracia, Enrique and López-Quílez, Antonio and Marco, Mar{\'i}a and Lila, Marisol},
  title   = {Neighborhood characteristics and violence behind closed doors: The spatial overlap of child maltreatment and intimate partner violence},
  journal = {PLOS ONE},
  year    = {2018},
  volume  = {13},
  number  = {6},
  pages   = {e0198684},
  doi     = {10.1371/journal.pone.0198684}
}

@article{Muldoon2019,
author = {Muldoon, K. A. and Galway, L. and Reeves, A. and Leach, T. and Heimerl, M. and Sampsel, K.},
title = {Geographies of Sexual Assault: A Spatial Analyses to Identify Neighborhoods Affected by Sexual and Gender-Based Violence},
journal = {Journal of Interpersonal Violence},
volume = {36},
number = {17-18},
pages = {8817--8834},
year = {2021},
doi = {10.1177/0886260519851175}
}

@article{VicenteSER2020,
  author  = {Vicente, G. and Goicoa, T. and Ugarte, M. D.},
  title   = {Bayesian inference in multivariate spatio-temporal areal models using INLA: analysis of gender-based violence in small areas},
  journal = {Stochastic Environmental Research and Risk Assessment},
  volume  = {34},
  number  = {10},
  pages   = {1421--1440},
  year    = {2020},
  doi     = {10.1007/s00477-020-01808-x}
}

@article{VicenteJRSSA2020,
    author = {Vicente, G. and Goicoa, T. and Fernandez-Rasines, P. and Ugarte, M. D.},
    title = {Crime Against Women in India: Unveiling Spatial Patterns and Temporal Trends of Dowry Deaths in the Districts of Uttar Pradesh},
    journal = {Journal of the Royal Statistical Society Series A: Statistics in Society},
    volume = {183},
    number = {2},
    pages = {655-679},
    year = {2020},
    month = {02},
    issn = {0964-1998},
    doi = {10.1111/rssa.12545},
}

@article{Habyarimana2021,
author = {Faustin Habyarimana and Temesgen Zewotir and Shaun Ramroop},
title ={Structured Spatial Modeling and Mapping of Domestic Violence Against Women of Reproductive Age in Rwanda},

journal = {Journal of Interpersonal Violence},
volume = {36},
number = {5-6},
pages = {2430-2454},
year = {2021},
doi = {10.1177/0886260518757222},
    note ={PMID: 29502504}
}

@article{Camargo2022,
  author  = {Camargo-Freile, Iv{\'a}n E. and Fl{\'o}rez-Lozano, Karen C. and Sarmiento-Crespo, Carlos A. and others},
  title   = {Risk of violence from a current or former partner: Associated factors and classification in a nationwide study in Colombia},
  journal = {PLOS ONE},
  year    = {2022},
  volume  = {17},
  number  = {12},
  pages   = {e0279444},
  doi     = {10.1371/journal.pone.0279444}
}

@article{GelfandVounatsou2003,
    author = {Gelfand, Alan E. and Vounatsou, Penelope},
    title = {Proper multivariate conditional autoregressive models for spatial data analysis},
    journal = {Biostatistics},
    volume = {4},
    number = {1},
    pages = {11-15},
    year = {2003},
    month = {01},
    issn = {1465-4644},
    doi = {10.1093/biostatistics/4.1.11},
}

@article{MartinezBeneito2013,
 ISSN = {00063444},
 URL = {http://www.jstor.org/stable/43305522},
 author = {Miguel A. Martinez-Beneito},
 journal = {Biometrika},
 number = {3},
 pages = {539--553},
 publisher = {[Oxford University Press, Biometrika Trust]},
 title = {A general modelling framework for multivariate disease mapping},
 urldate = {2026-08-14},
 volume = {100},
 year = {2013}
}

@techreport{INSProtocolo,
  author       = {{Instituto Nacional de Salud}},
  title        = {Protocolo de Vigilancia en Salud Pública: Violencia de género e intrafamiliar y ataques con agentes químicos},
  year         = {2023},
  institution  = {Instituto Nacional de Salud},
  address      = {Bogotá, Colombia},
  url          = {https://www.ins.gov.co/buscador-eventos/Lineamientos/Pro_Violencia%20de%20g%C3%A9nero.pdf}
}

@techreport{INS875,
  author       = {{Instituto Nacional de Salud}},
  title        = {Ficha de Notificación Individual: Violencia intrafamiliar y de género (INS 875)},
  year         = {2023},
  institution  = {Instituto Nacional de Salud},
  address      = {Bogotá, Colombia},
  url          = {https://www.ins.gov.co/Direcciones/Vigilancia/sivigila/FichasdeNotificacion/VIOLENCIA%20INTRAFAMILIAR%20F875.pdf}
}

@article{Rue2009,
author = {Rue, Håvard and Martino, Sara and Chopin, Nicolas},
title = {Approximate Bayesian inference for latent Gaussian models by using integrated nested Laplace approximations},
journal = {Journal of the Royal Statistical Society: Series B (Statistical Methodology)},
volume = {71},
number = {2},
pages = {319-392},
doi = {10.1111/j.1467-9868.2008.00700.x},
year = {2009}
}

@manual{INLApackage,
  title  = {R-INLA: Approximate Bayesian inference using Integrated Nested Laplace Approximations},
  author = {{INLA Team}},
  year   = {2024},
  url    = {https://www.r-inla.org}
}

@techreport{INSInforme2023,
  author       = {{Instituto Nacional de Salud}},
  title        = {Violencia de género e intrafamiliar. Informe de evento, Colombia, 2023},
  year         = {2023},
  institution  = {Instituto Nacional de Salud},
  address      = {Bogotá, Colombia},
  url = {https://www.ins.gov.co/buscador-eventos/Informesdeevento/VIOLENCIA%20DE%20GENERO%20INFORME%20DE%20EVENTO%202023.pdf}
}

@article{Yodanis2004,
author = {Carrie L. Yodanis},
title ={Gender Inequality, Violence Against Women, and Fear: A Cross-National Test of the Feminist Theory of Violence Against Women},
journal = {Journal of Interpersonal Violence},
volume = {19},
number = {6},
pages = {655-675},
year = {2004},
doi = {10.1177/0886260504263868},
    note ={PMID: 15140317}
}

@book{Young2005,
    author = {Young, Iris Marion},
    title = {On Female Body Experience: “Throwing Like a Girl” and Other Essays},
    publisher = {Oxford University Press},
    year = {2005},
    month = {01},
    isbn = {9780195161922},
    doi = {10.1093/0195161920.001.0001},
}

@article{Piedalueetal2020,
author = {Amy Piedalue and Amanda Gilbertson and Kalissa Alexeyeff and Elise Klein},
title ={Is Gender-Based Violence a Social Norm? Rethinking Power in a Popular Development Intervention},

journal = {Feminist Review},
volume = {126},
number = {1},
pages = {89-105},
year = {2020},
doi = {10.1177/0141778920944463}
}

@article{GouldAgnich2016,
author = {Laurie A. Gould and Laura E. Agnich},
title ={Exploring the Relationship Between Gender Violence and State Failure: A Cross-National Comparison},

journal = {Violence Against Women},
volume = {22},
number = {11},
pages = {1343-1370},
year = {2016},
doi = {10.1177/1077801215624790},
    note ={PMID: 26834148}
}

@article{KnorrHeldBest2001,
 ISSN = {09641998, 1467985X},
 URL = {http://www.jstor.org/stable/2680535},
 author = {Leonhard Knorr-Held and Nicola G. Best},
 journal = {Journal of the Royal Statistical Society. Series A (Statistics in Society)},
 number = {1},
 pages = {73--85},
 publisher = {[Oxford University Press, Royal Statistical Society]},
 title = {A Shared Component Model for Detecting Joint and Selective Clustering of Two Diseases},
 urldate = {2026-08-11},
 volume = {164},
 year = {2001}
}

@article{Held2005,
author = {Leonhard Held and Isabel Natário and Sarah Elaine Fenton and Håvard Rue and Nikolaus Becker},
title = {Towards joint disease mapping},
journal = {Statistical Methods in Medical Research},
volume = {14},
number = {1},
pages = {61-82},
year = {2005},
doi = {10.1191/0962280205sm389oa},
note = {PMID: 15691000}
}

@article{KnorrHeld2000,
author = {Knorr-Held, Leonhard},
title = {Bayesian modelling of inseparable space-time variation in disease risk},
journal = {Statistics in Medicine},
volume = {19},
number = {17-18},
pages = {2555-2567},
url = {https://onlinelibrary.wiley.com/doi/abs/10.1002/1097-0258%2820000915/30%2919%3A17/18%3C2555%3A%3AAID-SIM587%3E3.0.CO%3B2-%23},
year = {2000}
}

@article{Goicoa2018,
author = {Goicoa, T. and Adin, A. and Ugarte, M. D. and Hodges, J. S.},
title = {In spatio-temporal disease mapping models, identifiability constraints affect PQL and INLA results},
journal = {Stochastic Environmental Research and Risk Assessment},
volume = {32},
number = {3},
pages = {749--770},
year = {2018},
doi = {10.1007/s00477-017-1405-0}
}

@article{Retegui2024,
  author  = {Retegui, G. and Etxeberria, J. and Ugarte, M. D.},
  title   = {Multivariate Bayesian models with flexible shared interactions for analyzing spatio-temporal patterns of rare cancers},
  journal = {Environmental and Ecological Statistics},
  volume  = {31},
  pages   = {977--1007},
  year    = {2024},
  doi     = {10.1007/s10651-024-00630-w}
}

@article{simpson2017pc,
  author    = {Simpson, Daniel and Rue, H{\aa}vard and Riebler, Andrea and Martins, Thiago G. and S{\o}rbye, Sigrunn H.},
  title     = {Penalising Model Component Complexity: A Principled, Practical Approach to Constructing Priors},
  journal   = {Statistical Science},
  volume    = {32},
  number    = {1},
  pages     = {1--28},
  year      = {2017},
  month     = feb,
  doi       = {10.1214/16-STS576}
}

@article{Adin2023,
author = {Aritz Adin and Tomás Goicoa and James S Hodges and Patrick M Schnell and María D Ugarte},
title ={Alleviating confounding in spatio-temporal areal models with an application on crimes against women in India},

journal = {Statistical Modelling},
volume = {23},
number = {1},
pages = {9-30},
year = {2023},
doi = {10.1177/1471082X211015452}
}

@article{Piquero2021,
title = {Domestic violence during the COVID-19 pandemic - Evidence from a systematic review and meta-analysis},
journal = {Journal of Criminal Justice},
volume = {74},
pages = {101806},
year = {2021},
issn = {0047-2352},
doi = {10.1016/j.jcrimjus.2021.101806},
author = {Alex R. Piquero and Wesley G. Jennings and Erin Jemison and Catherine Kaukinen and Felicia Marie Knaul}
}

@article{Evans2020,
  author  = {Evans, Megan L. and Lindauer, Margo and Farrell, Maureen E.},
  title   = {A Pandemic within a Pandemic --- Intimate Partner Violence during Covid-19},
  journal = {New England Journal of Medicine},
  volume  = {383},
  number  = {24},
  pages   = {2302--2304},
  year    = {2020},
  month   = sep,
  doi     = {10.1056/NEJMp2024046}
}

@article{spiegelhalter2002,
author = {Spiegelhalter, David J. and Best, Nicola G. and Carlin, Bradley P. and Van Der Linde, Angelika},
title = {Bayesian measures of model complexity and fit},
journal = {Journal of the Royal Statistical Society: Series B (Statistical Methodology)},
volume = {64},
number = {4},
pages = {583-639},
doi = {10.1111/1467-9868.00353},
year = {2002}
}

@article{watanabe2010,
author = {Watanabe, Sumio},
title = {Asymptotic Equivalence of Bayes Cross Validation and Widely Applicable Information Criterion in Singular Learning Theory},
year = {2010},
issue_date = {3/1/2010},
publisher = {JMLR.org},
volume = {11},
issn = {1532-4435},
journal = {J. Mach. Learn. Res.},
month = dec,
pages = {3571--3594},
numpages = {24}
}

@techreport{world_health_organization_violence_2023,
  author      = {{World Health Organization}},
  title       = {Violence against women prevalence estimates, 2023},
  institution = {World Health Organization},
  address     = {Geneva},
  year        = {2025},
  isbn        = {978-92-4-011696-2},
  url         = {https://www.who.int/publications/i/item/9789240116962}
}

@article{MOYAGOMEZ2021,
title = {Exploring night and day socio-spatial segregation based on mobile phone data: The case of Medellin (Colombia)},
journal = {Computers, Environment and Urban Systems},
volume = {89},
pages = {101675},
year = {2021},
issn = {0198-9715},
doi = {10.1016/j.compenvurbsys.2021.101675},
author = {Borja Moya-Gómez and Marcin Stępniak and Juan Carlos García-Palomares and Enrique Frías-Martínez and Javier Gutiérrez}
}

@techreport{MedellinProjections,
  author      = {{Alcald\'ia de Medell\'in, Departamento Administrativo de Planeaci\'on}},
  title       = {Proyecciones de poblaci\'on, viviendas y hogares},
  institution = {Subdirecci\'on de Prospectiva, Informaci\'on y Evaluaci\'on Estrat\'egica},
  address     = {Medell\'in, Colombia},
  year        = {2024},
  url         = {https://www.medellin.gov.co/es/centro-documental/proyecciones-poblacion-viviendas-y-hogares/}
}

@misc{MENSchoolSuspension2020,
  author = {{Ministerio de Educación Nacional}},
  title = {Gobierno Nacional anuncia medidas en materia educativa
           para ofrecer garantías de salud pública a la comunidad},
  year = {2020},
  month = mar,
  note = {15 March 2020},
  url = {https://www.mineducacion.gov.co/1780/w3-article-393933.html?_noredirect=1}
}

@misc{MedellinAlternancia2021,
  author = {{Alcaldía de Medellín}},
  title = {Este miércoles inicia la alternancia en las instituciones
           educativas oficiales de {Medellín}},
  year = {2021},
  month = feb,
  note = {8 February 2021},
  url = {https://www.medellin.gov.co/es/?p=35656}
}

@misc{MedellinReactivacion2021,
  author = {{Alcaldía de Medellín}},
  title = {{Medellín} reactiva el modelo de alternancia educativa
           a partir del próximo martes 18 de mayo},
  year = {2021},
  month = may,
  note = {14 May 2021},
  url = {https://www.medellin.gov.co/es/sala-de-prensa/noticias/medellin-reactiva-el-modelo-de-alternancia-educativa-a-partir-del-proximo-martes-18-de-mayo/}
}

@misc{MedellinPresencialidad2021,
  author = {{Alcaldía de Medellín}},
  title = {310.000 estudiantes de las instituciones educativas
           oficiales de {Medellín} retoman las clases},
  year = {2021},
  month = jul,
  url = {https://www.medellin.gov.co/es/sala-de-prensa/noticias/310-000-estudiantes-de-las-instituciones-educativas-oficiales-de-medellin-retoman-las-clases/}
}

@article{Baker2017,
  author  = {Baker, Jannah and White, Nicole and Mengersen, Kerrie
             and Rolfe, Margaret and Morgan, Geoffrey G.},
  title   = {Joint modelling of potentially avoidable hospitalisation
             for five diseases accounting for spatiotemporal effects:
             A case study in New South Wales, Australia},
  journal = {PLOS ONE},
  year    = {2017},
  volume  = {12},
  number  = {8},
  pages   = {e0183653},
  doi     = {10.1371/journal.pone.0183653}
}

@article{gelman2014,
  author  = {Gelman, Andrew and Hwang, Jessica and Vehtari, Aki},
  title   = {Understanding predictive information criteria for {B}ayesian models},
  journal = {Statistics and Computing},
  volume  = {24},
  number  = {6},
  pages   = {997--1016},
  year    = {2014},
  doi     = {10.1007/s11222-013-9416-2}
}

@article{vehtari2017,
  author  = {Vehtari, Aki and Gelman, Andrew and Gabry, Jonah},
  title   = {Practical {B}ayesian model evaluation using leave-one-out cross-validation and {WAIC}},
  journal = {Statistics and Computing},
  volume  = {27},
  number  = {5},
  pages   = {1413--1432},
  year    = {2017},
  doi     = {10.1007/s11222-016-9696-4}
}

@article{czado2009,
  author  = {Czado, Claudia and Gneiting, Tilmann and Held, Leonhard},
  title   = {Predictive model assessment for count data},
  journal = {Biometrics},
  volume  = {65},
  number  = {4},
  pages   = {1254--1261},
  year    = {2009},
  doi     = {10.1111/j.1541-0420.2009.01191.x}
}

@misc{MedellinIMCV,
  author       = {{Alcald\'ia de Medell\'in, Departamento Administrativo de Planeaci\'on}},
  title        = {\'Indice Multidimensional de Condiciones de Vida ({IMCV}) por comuna y corregimiento, 2011--2018},
  howpublished = {MEData, Medell\'in open-data platform},
  url          = {},
  note         = {Accessed ...}
}

@Article{Sepulveda2019,
journal={Social Indicators Research: An International and Interdisciplinary Journal for Quality-of-Life Measurement},
author={Fabio Humberto SepÃºlveda Murillo and Jorge Chica Olmo and Norely Margarita Soto Builes},
title={Spatial Variability Analysis of Quality of Life and Its Determinants: A Case Study of MedellÃ­n, Colombia},
year={2019},
month={August},
pages={1233-1256},
volume={144},
number={3},
doi={10.1007/s11205-019-02088-x},
url={https://ideas.repec.org/a/spr/soinre/v144y2019i3d10.1007_s11205-019-02088-x.html},
}

@misc{MedellinCartography,
  author       = {{Alcald\'ia de Medell\'in}},
  title        = {{Barrios y Veredas}},
  howpublished = {GeoMedell\'in, cartographic layer},
  year         = {n.d.},
  url          = {https://www.medellin.gov.co/servidormapas/rest/services/mapas_nacionales/VC_Limite_Politico_Admtivo/MapServer/0},
  note         = {Accessed 11 October 2026}
}

@misc{MedellinReopening2021,
  author = {{Alcald\'ia de Medell\'in}},
  title  = {{Medell\'in} espera recibir 250.000 turistas y una
            derrama econ\'omica por m\'as de \$178.000 millones
            durante la temporada decembrina},
  year   = {2021},
  month  = dec,
  note   = {2 December 2021},
  url    = {https://www.medellin.gov.co/es/sala-de-prensa/noticias/medellin-espera-recibir-250-000-turistas-y-una-derrama-economica-por-mas-de-178-000-millones-durante-la-temporada-decembrina/}
}

@misc{MedellinPresencialidad2022,
  author = {{Alcald\'ia de Medell\'in}},
  title  = {Las instituciones educativas oficiales de {Medell\'in}
            continuar\'an con clases 100~\% presenciales durante 2022},
  year   = {2022},
  month  = jan,
  note   = {6 January 2022},
  url    = {https://www.medellin.gov.co/es/sala-de-prensa/noticias/las-instituciones-educativas-oficiales-de-medellin-continuaran-con-clases-100-presenciales-durante-2022/}
}

@misc{WHOVAWFactsheet,
  author       = {{World Health Organization}},
  title        = {Violence against women},
  year         = {2026},
  howpublished = {Fact sheet},
  url          = {https://www.who.int/news-room/fact-sheets/detail/violence-against-women},
  note         = {Updated 18 June 2026; accessed 12 October 2026}
}

@article{Valganon2023,
  author  = {Valga{\~n}{\'o}n, Pablo
             and Useche, Andr{\'e}s F.
             and Soriano-Pa{\~n}os, David
             and Ghoshal, Gourab
             and G{\'o}mez-Garde{\~n}es, Jes{\'u}s},
  title   = {Quantifying the heterogeneous impact of lockdown policies
             on different socioeconomic classes during the first
             {COVID-19} wave in {Colombia}},
  journal = {Scientific Reports},
  year    = {2023},
  volume  = {13},
  pages   = {16481},
  doi     = {10.1038/s41598-023-43685-8}
}

@misc{MEDataViolence,
  author       = {{Alcald\'ia de Medell\'in}},
  title        = {Vigilancia Integrada de la Violencia de G\'enero},
  year         = {2023},
  howpublished = {Secretar\'ia de Salud. {MEData}, dataset},
  url          = {https://medata.app.medellin.gov.co/node/16578},
  note         = {Last updated 20 September 2023; accessed 12 October 2026}
}

@techreport{MedellinCandelariaPlan,
  author      = {{Alcald\'ia de Medell\'in}},
  title       = {Plan Local de Seguridad y Convivencia:
                 Comuna 10 -- La Candelaria},
  institution = {Secretar\'ia de Seguridad y Convivencia},
  address     = {Medell\'in, Colombia},
  year        = {n.d.},
  url         = {https://www.medellin.gov.co/es/wp-content/uploads/2022/07/Plan-local-de-seguridad-y-convivencia-Comuna-10_Candelaria.pdf},
  note        = {Accessed 12 October 2026}
}

@misc{WHOSexualViolenceSurveillance,
  author       = {{World Health Organization}},
  title        = {Violence against women: non-partner sexual violence
                  prevalence among women in the previous 12 months (\%)},
  year         = {n.d.},
  howpublished = {Global Health Observatory, indicator metadata},
  url          = {https://www.who.int/data/gho/data/indicators/indicator-details/GHO/non-partner-sexual-violence-prevalence-%28-%29},
  note         = {Accessed 12 October 2026}
}

\end{document}